\documentclass{aa}  

\usepackage{graphicx}
\usepackage{xcolor}
\usepackage{txfonts}
\usepackage[]{hyperref}
\usepackage{CJKutf8}

\newcommand{\totalstarnumber}{2495}
\newcommand{\doubletnumber}{800}
\newcommand{\tripletnumber}{1206}
\newcommand{\rotationnumber}{2006}
\newcommand{\norotationnumber}{489}
\newcommand{\threesigmapositiverotationnumber}{243}
\newcommand{\positiveenveloperotationstarnumber}{243}

\newcommand{\vect}[1]{\boldsymbol{\rm #1}}
\newcommand{\seb}[1]{{#1}}
\newcommand{\sebnew}[1]{\textcolor{black}{#1}}
\newcommand{\jb}[1]{{#1}}
\newcommand{\jbnew}[1]{\textcolor{black}{#1}}
\newcommand{\LG}[1]{{#1}}
\newcommand{\LGnew}[1]{\textcolor{black}{#1}}
\newcommand{\CNnames}[1]{{\begin{CJK}{UTF8}{gbsn}~(#1)~\end{CJK}}}

\newcommand{\LGvtwo}[1]{\textcolor{black}{#1}}

\begin{document}

   \title{Asteroseismic measurement of core and envelope rotation rates for \rotationnumber~red giant branch stars}

   %\subtitle{}

   \author{Gang Li \CNnames{李刚}
          \inst{1,2}\fnmsep\thanks{\LGvtwo{The electronic table is available at the Zenodo link...(I would like to update this part when the paper is officially accepted and uploaded to ArXiv, because I think it is better to provide a DOI or an ArXiv number to Zenodo.)} }
          \and
          S\'{e}bastien Deheuvels\inst{1}
          \and
          J\'{e}r\^ome Ballot\inst{1}
          }

   \institute{IRAP, Université de Toulouse, CNRS, CNES, UPS, 14 avenue Edouard Belin, 31400 Toulouse, France\\
              \email{gang.li@kuleuven.be, sebastien.deheuvels@irap.omp.eu}
         \and
         Institute of Astronomy, KU Leuven, Celestijnenlaan 200D, 3001 Leuven, Belgium.
         %    University of Alexandria, Department of Geography, ...\\
          %   \email{c.ptolemy@hipparch.uheaven.space}
         %    \thanks{The university of heaven temporarily does not
          %           accept e-mails}
             }

   \date{Received September 15, 1996; accepted March 16, 1997}

% \abstract{}{}{}{}{} 
% 5 {} token are mandatory
 
  \abstract
  % context heading (optional)
  % {} leave it empty if necessary  
   {Tens of thousands of red giant stars in the \emph{Kepler} data exhibit solar-like oscillations. The mixed-mode characteristics of their oscillations enable us to study the internal physics from core to surface, such as differential rotation. However, envelope rotation rates have been measured for only a dozen red-giant-branch (RGB) stars so far. The limited sample hinders the theoretical interpretation of angular momentum transport in post-main-sequence phases.
   }
   {We report the measurements of g-mode properties and differential rotation in the largest sample of \emph{Kepler} RGB stars.}
   {We apply a new approach to calculate the asymptotic frequencies of mixed modes, which accounts for the so-called near-degeneracy effects and leads to more proper measurements of envelope rotation rates. By fitting these asymptotic expressions to the observations, we obtain measurements of the properties of g modes (period spacing $\Delta \Pi_1$, coupling factor $q$, g-mode offset term $\varepsilon_g$, small separation $\delta \nu_{01}$) and the internal rotation (mean core $\Omega_\mathrm{core}$ and envelope $\Omega_\mathrm{env}$ rotation rates). 
    }
   {
   Among \totalstarnumber~stars with clear mixed-mode patterns, we found that \doubletnumber~show doublets, \tripletnumber~show triplets, while the remaining stars do not show any rotational splittings. 
   We measured core rotation rates for \rotationnumber~red giants, doubling the size of pre-existing catalogues. This led us to discover an over-density of stars that narrowly distribute around a well-defined ridge in the plane showing core rotation rate versus evolution along the RGB. These stars could experience a different angular momentum transport compared to other red giants. With this work, we also increase the sample of stars with measured envelope rotation rates by two orders of magnitude. We find a decreasing trend between envelope rotation rates and evolution, 
   implying that the envelopes slow down with expansion, as expected. 
   We find \positiveenveloperotationstarnumber~stars whose envelope rotation rates are significantly larger than zero. For these stars, the core-to-envelope rotation ratios are around $\Omega_\mathrm{core}/\Omega_\mathrm{env}\sim20$ and show a large spread with evolution. Several stars show extremely mild differential rotations, with core-to-surface ratios between 1 and 2. These stars also have very slow core rotation rates, suggesting that they go through a peculiar rotational evolution. We also discovered more stars located below the $\Delta \Pi_1$ -- $\Delta \nu$ degeneracy sequence, which will provide the opportunity to study the history of possible stellar mergers. 
   }
  % conclusions heading (optional), leave it empty if necessary 
   {}


   \maketitle
%
%------------------------------------------------------------------- 
\section{Introduction \label{sec:introduction}}
%Some perturbation mechanisms within stars can induce oscillations, resulting in periodic expansions and contractions of their structures. These oscillations subsequently manifest as variations in stellar brightness \citep[e.g.][]{Aerts2010}. 

Thanks to the advent of recent space-based photometric missions, particularly the \emph{Kepler} mission \citep{Borucki2010Sci}, stellar oscillations were observed with a level of precision that was previously unimaginable through ground-based observations. One remarkable achievement in the field of asteroseismic research is related to red giants, which are stars that have depleted their central hydrogen and are now burning hydrogen in a shell. The oscillations in red giants are stochastically excited in the outer convective envelope, which is similar to the Sun. To date, tens of thousands of solar-like oscillators have been discovered by \emph{Kepler} \citep{Bedding2010ApJ, Yu2018}. 

The power spectra of solar-like oscillators exhibit distinct regular-spaced radial ($l=0$) and quadrupole ($l=2$) modes, where $l$ represents the angular degree. In post-main-sequence solar-like stars, dipole ($l=1$) modes show a mixed-mode nature. This arises from the coupling between the outer pressure modes and the interior gravity modes \citep{Bedding2014}. Unlike the equally-spaced frequencies or periods predicted by the asymptotic relation \citep{shibahashi79, Tassoul1980ApJS}, the frequencies of mixed modes display a more complex pattern \citep[e.g.,][]{Goupil13, mosser12a, Mosser2015}. The coupling between the stellar interior and outer envelope allows us to investigate the physics from the stellar core to the surface. For example, mixed modes have been used to distinguish between Hydrogen-shell-burning and Helium-core-burning giants \citep{Bedding2011} \LGvtwo{and} unveil indications of mass transfer or merger \citep{Deheuvels2022A&A, Rui2021MNRAS, Tayar2022, LiYaGuang2022NatAs}. \LGvtwo{Mixed modes are predicted to reveal central magnetic fields, as these fields lead to frequency shifts \citep{Gomes2020, Loi2021, Bugnet2021, Mathis2021, Li2022Nature, Mathis2023}, modifications in period spacing \citep{Loi2020, Li2022Nature, Bugnet2022, Deheuvels2023}, and amplitude suppression \citep{Fuller2015, Stello2016, Mosser2017_suppressed_mode, Loi2017, Loi2018}. The intensities and topology of central magnetic fields in dozens of stars have been reported by \cite{Li2022Nature}, \cite{Deheuvels2023}, and \cite{Li2023}. }

\LGvtwo{Pertinent to the interests of this work, mixed modes can also be used to measure the internal rotation rates of red giant stars, most of which are core rotation rates \citep{Beck2012Natur, Gehan2018, Kuszlewicz2023ApJ}. As a contrast, the envelope rotation rates have only been reported for dozens of stars \citep{deheuvels12, Deheuvels2014, DiMauro2016, Triana2017}.}

Stellar rotation plays a crucial role in the process of stellar evolution \citep{Maeder2009}. It leads to the distortion of stellar structure and facilitates the transport of additional hydrogen to the core, thereby extending the star's lifetime. Asteroseismology offers a powerful method for accurately measuring rotation rates in different depths of a star. 
Studies have revealed that main-sequence stars, not only Sun-like (including the Sun), but also hotter $\gamma$\,Doradus stars (A to F type), show nearly uniform rotations \citep{Schou1998, Gough2015, Benomar2015, Nielsen2015, LiGang2020, Aerts2019ARA&A}. The rotation periods of main-sequence stars decrease with increasing masses, ranging from tens of days to one day, consistent with spectroscopic observations \citep{Royer2007, McQuillan2014, VanReeth2016, LiGang2020, Saio2021}. 
After the main sequence, stars tend to rotate differentially. \cite{Deheuvels2014, Deheuvels2020} found a clear transition from near-uniform to differential rotations in young subgiant stars. After that, the core rotation rates of red giants show no correlation with their evolution and mass, even though their cores are contracting, which implies that some additional and efficient angular momentum transport mechanisms are at work \citep{Mosser2012,Wu2016ApJ, Gehan2018}. As stars evolve beyond the red-giant-branch (RGB) stage and enter the helium-burning phases, the amount of differential rotation decreases \citep{Deheuvels2015}.

Despite its recognised importance, the theoretical understanding of angular momentum transport remains a primary challenge in current stellar physics. Firstly, rotation rates are consistently slower than predicted at nearly all stages of stellar evolution \citep{Ouazzani2019, LiGang2020, Fuller2019}. Secondly, the pure hydrodynamic processes that involve angular momentum transfer through meridional circulation and shear instabilities \citep[e.g.][]{Zahn1992, Mathis2004, Mathis2018} generate excessively strong differential rotation, which is not supported by observations \citep[e.g.][]{Eggenberger2012, Ceillier2013, Marques2013}. These discrepancies imply the existence of unknown mechanisms responsible for internal angular momentum transport within stars. \seb{The efficiency of these mechanisms was found to increase with stellar mass and evolution on the red giant branch \citep{Cantiello2014, Eggenberger2015IAUS, Spada2016}. By contrast, it was shown that the efficiency of angular momentum transport decreases with evolution during the subgiant phase \citep{Eggenberger2019I}.}
%When a star ascends the red giant branch, the effect of unknown mechanisms was found to increase with mass and evolutionary stages to reproduce the observations \citep{Cantiello2014, Eggenberger2015IAUS, Spada2016}. While during the subgiant phase, the effect increases with stellar mass but decreases with evolution \citep{Eggenberger2019I}. 

Several mechanisms have been proposed to reconcile the discrepancy between theoretical predictions and observations. One such mechanism involves internal gravity waves (IGW) that are excited at the base of the convective envelope and contribute to additional angular momentum transport \citep{Fuller2014, Pincon2016}. However, IGWs still face challenges in explaining the rotational behaviour of more evolved red giants. Mixed modes, on the other hand, 
%have been considered as a potential mechanism 
\seb{could be efficient at transporting angular momentum on the upper RGB} 
\citep{Belkacem2015}. Recently, 
%magnetic fields in the vicinity of hydrogen-burning shells 
core magnetic fields were discovered in some red giant stars \citep{Li2022Nature, Deheuvels2023, Li2023}, providing observational constraints on how magnetic fields influence angular momentum transport \citep[e.g.][]{Maeder2014ApJ, Kissin2015ApJ, Rudiger2015A&A, Jouve2015, Fuller2019, Barrere2023, Petitdemange2023, Petitdemange2024}. 
\LGvtwo{Recently, \cite{Eggenberger2022NatAs} showed the importance of the Tayler instability in the Sun to reproduce the observed internal rotation profile and lithium and helium abundances. Angular momentum transport via the Tayler instability appears to better match the asteroseismic core rotation rates of both main-sequence and post-main-sequence stars \citep{Moyano2023,Eggenberger2022}. }
However, theories based on magnetism still struggle to fully explain the observed internal rotations. Another important mechanism to consider is binary interaction. It has been demonstrated that tidal effects from binary companions can significantly shape the internal rotations of stars \citep{LiGang2020binary, Fuller2021, Ahuir2021}. Given the prevalence of binary systems, the tidal effects they induce cannot be ignored and may have a substantial impact on stellar rotation.

Current studies typically report only the core rotation rates of RGB %red giant branch (RGB) 
stars, operating under the assumption that the envelope rotation rates are negligible and can be treated as zero. There are samples of roughly one thousand stars whose core rotation rates have been measured \citep{Gehan2018, Kuszlewicz2023ApJ}. Conversely, a limited number of red giants, approximately %a dozen
\seb{twenty}, have had their envelope rotation rates directly measured \citep{Deheuvels2014,DiMauro2016, Triana2017}. \seb{This type of measurement is complicated for several reasons. First, the strong expansion of red-giant envelopes causes them to spin down, so the envelope rotation periods are long and become comparable to the duration of \textit{Kepler} observations for more evolved red giants. Secondly, while rotation inversions are very efficient at providing estimates of core rotation rates, they can lead to biased measurements for the envelope rotation rates because of the pollution from the core (\citealt{Deheuvels2014}). This is all the more true for evolved giants. Finally, as we show in the present paper, the measurements of envelope rotation rates can be strongly affected by the so-called near-degeneracy effects (NDE), which arise when the frequency separation between consecutive mixed modes is comparable to the rotational splitting (\citealt{Deheuvels2017}). None of the studies performed so far to measure the internal rotation of red giants using dipole mixed modes accounted for these NDE.}
%In our research, we treat the envelope rotation rates as a free parameter to investigate whether we can effectively constrain these rates using current data of solar-like mixed modes.

\jb{In the present work,
we have analysed existing seismic observations of red giant stars obtained by the \emph{Kepler} mission to measure both the core and envelope rotation rates for a large number of targets. We thus treated the envelope rotation rates as free parameters and investigated whether we can effectively constrain their values.}
%Additionally, 
We introduced a new fitting method based on the asymptotic expression of mixed modes, which can %reproduce 
\seb{properly account for NDE.} 
%the so-called near-degeneracy effects \citep[NDE,][]{Deheuvels2017}. \seb{We indeed found that this was crucial to adequately measure envelope rotation rates in red giants.}
%Therefore, our work, which significantly expands the sample size for core rotation rates by a factor of two and for envelope rotation rates by a factor of 100, is of great importance. 
\jb{Finally, we were able to significantly expand the sample of RGB stars with measured internal rotation: the sample size has been doubled for core rotation rates and multiplied by a factor of 100 for envelope rotation rates.}

The paper is organised as follows: In section~\ref{sec:Asymptotic_theories}, we introduce %our asymptotic theories and how we fit the observed spectra. 
\seb{asymptotic expressions of mixed modes, which are then fit to the observed spectra.} We propose a new method that avoids using the $\zeta$ function and we show that 
%the new method works better
\seb{it is better suited to measure envelope rotation rates} in section~\ref{subsec:Approach_comparison}. We compare our results with the previous literature in section~\ref{subsec:comparison} and show that our results are generally consistent and reliable. In section \ref{sec:discussions}, 
%we give the discussions about our results, 
\seb{we present our results and comment on the distributions that we obtain for core and envelope rotation rates, and their dependence on stellar parameters. Special care is given to}
%including the new 
stars that might undergo a process of mass transfer or merger.
%, and differential rotations as functions of multiple parameters and evolution. 
Finally, the conclusions are given in section~\ref{sec:conclusion}.

\section{Data analysis}\label{sec:Asymptotic_theories}

\subsection{Sample collection}
\begin{figure}
    \centering
    \includegraphics[width=1\linewidth]{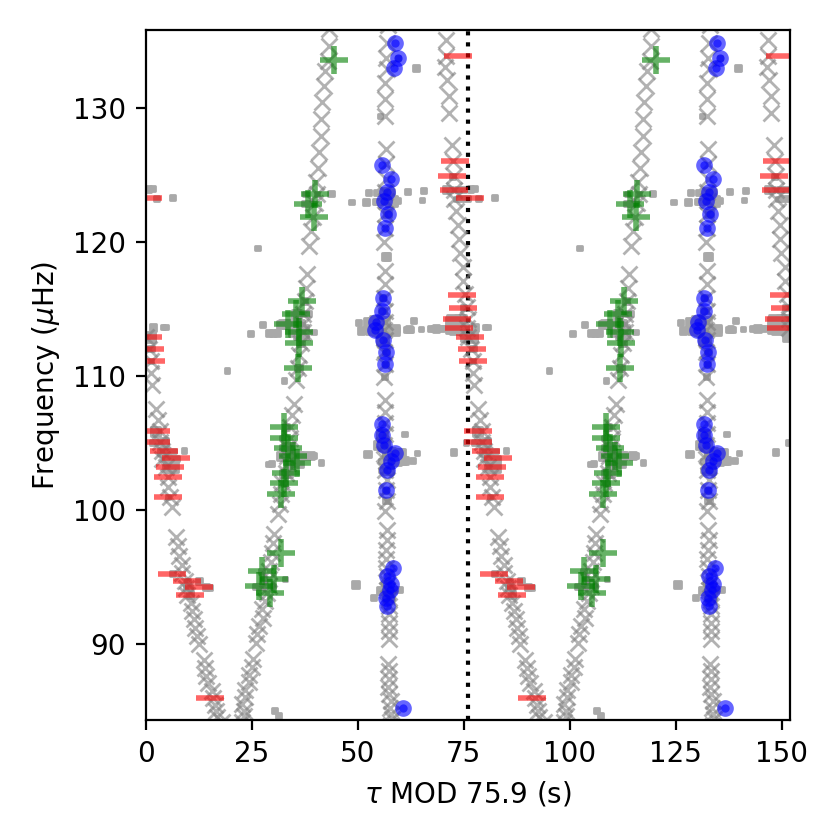}
    \caption{Stretched \'{e}chelle diagram of KIC\,8636389. The x-axis is the stretched period $\tau$ module 75.9 s. The y-axis is the oscillation frequency. The grey dots are the peaks with signal-to-noise ratio larger than 10. The blue circles are identified as $l=1$ $m=0$ mixed modes. The red line and green plus are $m=-1$ and $m=+1$ modes. The crosses show the best-fitting frequencies. }
    \label{fig:stretched_8636389}
\end{figure}

We collected the red giant stars from the catalogues by \cite{Yu2018} and \cite{Gehan2018} and used the \emph{Kepler} four-year long-cadence data downloaded from the Mikulski Archive for Space Telescopes (MAST) to perform the seismic analysis. We derived the frequency of the maximum power ($\nu_\mathrm{max}$), the initial estimate of the large frequency separation ($\Delta\nu$), and the Harvey background \citep{Karoff2008PhDT} by applying the algorithm used in the SYD pipeline \citep{Huber2009, Chontos2021}. 

\subsection{Asymptotic properties of pressure modes}

Before fitting the $l=1$ mixed modes, we needed to obtain information on %$l=0$ and $l=2$ p modes. 
the asymptotic properties of p modes. \seb{Quadrupole modes are in principle mixed modes. However, since the coupling between the p- and g-mode cavities is much weaker for these modes that for dipole modes, only $l=2$ modes that behave predominantly as p modes are visible.}
The frequencies of $l=0$ and $l=2$ modes ($\nu_\mathrm{p}$) are derived by fitting the oscillation spectrum with the Lorentzian profiles, which are written as
\begin{equation}
    P(\nu) = \frac{1}{4}\frac{a_0^2}{(\nu-\nu_0)^2+\eta^2},
\end{equation}
where $P(\nu)$ is the oscillation power, $a_0$ is the amplitude parameter, $\nu_0$ is the centre of the profile, and $\eta$ is the half width at half maximum. For each star, we ran a Markov chain Monte Carlo (MCMC) optimisation algorithm to search for the best-fitting results using the emcee package \citep{Foreman-Mackey2013PASP}. In the MCMC algorithm, we maximised the likelihood function, defined by \cite{Anderson1990ApJ} as
\begin{equation}
    \ln p\left(D|\theta, M \right) = -\sum_i\left[\ln M_i\left(\theta \right) +\frac{D_i}{M_i\left(\theta\right)} \right],
\end{equation}
where $D_i$ is the observed data, $\theta$ stands for the parameters, and $M_i$ is the model. In this MCMC algorithm, we obtained the best-fitting frequencies for $l=0$ and $l=2$ modes $\nu_{l, i}^\mathrm{obs}$ and their uncertainties $\sigma_{l, i}^\mathrm{obs}$.

\seb{For high-radial order p modes, the second-order asymptotic relation can be expressed as}
\begin{equation}
        \nu_\mathrm{p} = \left(n_\mathrm{p}+\frac{l}{2}+\varepsilon_\mathrm{p}+\frac{\alpha}{2}(n_\mathrm{p}-n_\mathrm{max})^2\right)\Delta\nu - l(l+1)D, \label{eq:p_mode_asymptotic_relation}
\end{equation}
where $\Delta \nu$ is the revised large separation, $\varepsilon_\mathrm{p}$ is the p-mode phase term, and $n_\mathrm{max}=\nu_\mathrm{max}/\Delta\nu$ is the radial order at the max-power frequency (\citealt{mosser12a}). The term $D$ describes the small separation between $l=0$ and $2$ modes \citep{shibahashi79,Tassoul1980ApJS,Mosser2011A&A}. 

We ran another MCMC optimisation algorithm to obtain the best-fitting $\Delta \nu$, $\varepsilon_\mathrm{p}$, $\alpha$, and $D$ by maximising the likelihood function $L_\mathrm{p}$ defined as:
\begin{equation}
    \ln L_\mathrm{p} = -\frac{1}{2} \sum_{l=0,2}\sum_{i} \left[ \frac{\left(\nu_{l, i}^\mathrm{obs}-\nu_{l, i}^\mathrm{cal}\right)^2}{\sigma_{l, i}^2} +\ln \left(2\pi\sigma_{l, i}^2\right) \right], \label{equ:likelihood_for_p_mode}
\end{equation}
where $\nu_{l, i}^\mathrm{obs}$ is the $i\mathrm{th}$ observed frequencies of $l=0$ and $l=2$ modes, $\nu_{l, i}^\mathrm{cal}$ is the theoretical frequencies calculated by Eq.~\ref{eq:p_mode_asymptotic_relation}, and $\sigma_{l, i}$ is the uncertainty. After fitting all the stars' p modes, we find that there is a residual spread with a median value % of $0.042\,\mathrm{\mu Hz}$.
\jb{ $\sigma_{\rm m} = 0.042\,\mathrm{\mu Hz}$}. 
%We consider it as the limit of the asymptotic relation of Eq.~\ref{eq:p_mode_asymptotic_relation}. 
\jb{We consider that this value $\sigma_{\rm m}$ is the model error due to the limitation of the asymptotic relation Eq.~\ref{eq:p_mode_asymptotic_relation}. }To properly account for it, the uncertainty of p-mode frequency is defined as a quadratic summation as
\begin{equation}
    %\sigma_{l, i}^2 = \sigma_{l, i}^\mathrm{obs}^2+(0.042\,\mathrm{\mu Hz})^2.
    \sigma_{l, i}^2 = \left(\sigma_{l, i}^\mathrm{obs}\right)^2+\jb{\sigma_{\rm m}^2.}
\end{equation}
In the MCMC code, we used 32 parallel chains and 2000 steps, retaining only the last 1600 steps as the final posterior distributions. The asymptotic expression of the p modes is simple and explicit, which allowed the MCMC code to complete and converge rapidly.

\subsection{Asymptotic expression for $l=1$ mixed modes}

After fitting and removing the $l=0$ and $l=2$ p modes, We followed the so-called `stretched echelle diagram' procedure introduced by \citep{Mosser2015} to extract the $l=1$ mixed modes and identify their azimuthal orders $m$. In this procedure, the unevenly-spaced $l=1$ mixed modes are stretched by a differential equation to generate vertical ridges in the stretched \'{e}chelle diagram. The vertical ridges provide the estimate of the period spacing $\Delta \Pi_1$, and the slight differences in %different 
\seb{the period spacings found for the}
ridges allow us to identify their azimuthal orders $m$. Figure~\ref{fig:stretched_8636389} shows a stretched \'{e}chelle diagram using KIC\,8636389 as an example.

To calculate the \seb{asymptotic frequencies of $l=1$ mixed modes (\citealt{shibahashi79}, \citealt{Unno1989}), we solve the equation}:
\begin{equation}
    \tan \theta_\mathrm{p} = q \tan \theta_\mathrm{g},\label{eq:asymptotic_expression}
\end{equation}
where $q$ is the coupling factor, which gives physical constraints on the regions surrounding the radiative core and the hydrogen-burning shell of subgiants and red giants \citep{Mosser2017}. $\theta_\mathrm{g}$ and $\theta_\mathrm{p}$ are phase terms characterising pure g and p modes, given by
\begin{equation}
    \theta_\mathrm{g} = \pi \frac{1}{\Delta\Pi_1}\left(\frac{1}{\nu}-\frac{1}{\nu_\mathrm{g}}\right),
\end{equation}
and
\begin{equation}
    \theta_\mathrm{p}=\pi\frac{\nu-\nu_{\mathrm{p}, l=1}}{\Delta\nu}.
    \label{eq_thetap}
\end{equation}

The frequencies $\nu_\mathrm{g}$ of pure $l=1$ g modes are equally spaced in period, and hence can be written as
\begin{equation}
    \frac{1}{\nu_\mathrm{g}}=P_\mathrm{g} = \Delta\Pi_1\left(n_\mathrm{g}+\varepsilon_\mathrm{g}\right), \label{eq:g_mode_equally_spacing_formula}
\end{equation}
where $\varepsilon_\mathrm{g}$ is %the phase of pure g modes. 
\seb{an offset term.}
\LG{Here, we \seb{adopt} the same definition of $\varepsilon_\mathrm{g}$ as \cite{Mosser2015}. However, as discussed by \cite{Lindsay2022ApJ} and \cite{Ong2023ApJ}, 
%we realised that our definition of $\varepsilon_\mathrm{g}$ in Eq.~\ref{eq:g_mode_equally_spacing_formula} 
\seb{with this definition,}
%cannot allow 
Eq.~\ref{eq:asymptotic_expression} \seb{fails} to recover 
%the situation 
\seb{the frequencies of pure g modes in the case } of a weak coupling (when $q \rightarrow 0$). We \seb{caution the reader} that there is a $\frac{1}{2}$ offset of $\varepsilon_\mathrm{g}$ in our work, so $\varepsilon_\mathrm{g}+\frac{1}{2}$ should be the correct value of $\varepsilon_\mathrm{g}$ using the proper definition.}

The terms $\nu_{\mathrm{p}, l=1}$ in Eq. \ref{eq_thetap} are the frequencies of $l=1$ pure p modes. They cannot be calculated directly from Eq.~\ref{eq:p_mode_asymptotic_relation} because the small separation $\delta\nu_\mathrm{01}$ \citep[the amount by which $l = 1$ modes are offset from the midpoint of consecutive $l = 0$ modes,][]{Bedding2014} \seb{varies} with evolution and cannot be reproduced well using only the $D$ parameter 
%in eq.~\ref{eq:p_mode_asymptotic_relation}
\citep{Lund2017ApJ}. Therefore, we \seb{include} an additional \seb{free} parameter $f_\mathrm{shift}$ to describe the shift of the $l=1$ pure p modes, as
\begin{equation}
        \nu_{\mathrm{p}, l=1} = \left(n_\mathrm{p}+\frac{1}{2}+\varepsilon_\mathrm{p}+\frac{\alpha}{2}(n_\mathrm{p}-n_\mathrm{max})^2\right)\Delta\nu - 2D + f_\mathrm{shift}. \label{eq:l_1_p_mode_asymptotic_relation}
\end{equation}

\LG{The parameter $f_\mathrm{shift}$ is related to the small separation $\delta\nu_{01} = \frac{1}{2}\left(  \nu_{n_\mathrm{p}, l=0} + \nu_{n_\mathrm{p}+1, l=0}  \right) - \nu_{n_\mathrm{p}, l=1}$. Ignoring the second-order term, we have $\delta \nu_{01} = 2D-f_\mathrm{shift}$.} We provide the discussion about the small separation $\delta \nu_{01}$ in appendix~\ref{appendix_sec:other_parameters}.
%Next, we introduced rotational perturbations to the mixed modes, as detailed in the following section.

\subsection{Rotational perturbation}

\seb{To take into account the rotational perturbations to mixed mode frequencies,} we attempted two different approaches
%to solve the eq.~\ref{eq:asymptotic_expression} and to calculate the splittings of mixed modes, 
and we will show that the second approach is a better way.

\subsubsection{Approach One}\label{subsec:approach1}

%The frequencies of $m=0$ modes $\nu_{m=0}$ are first obtained by solving Eq.~\ref{eq:asymptotic_expression}. 
\seb{Approach One is the commonly-used method used to calculate the rotational splittings $\delta \nu_\mathrm{rot}$ of mixed modes in red giants. It is now customary to follow \cite{Goupil13} and to express it as}
%Firstly, we solved eq.~\ref{eq:asymptotic_expression} to obtain the frequencies of $m=0$ modes $\nu_{m=0}$, and calculate the splittings $\delta \nu_\mathrm{rot}$ using the angular frequencies of the core rotation $\Omega_\mathrm{core}$ and the envelope rotation $\Omega_\mathrm{env}$, 
\begin{equation}
    2\pi \delta \nu_\mathrm{rot} = \frac{\Omega_\mathrm{core}}{2}\zeta + \Omega_\mathrm{env}\left(1-\zeta \right), \label{eq:splitting_vs_zeta}
\end{equation}
where $\Omega_\mathrm{core}$ is the average core rotation, $\Omega_\mathrm{env}$ is the average envelope rotation, and $\zeta$ is the ratio of mode inertia in the g cavity over the total mode inertia, hence $\zeta=1$ stands for a pure g mode while $\zeta=0$ is pure p modes \citep{deheuvels12, Goupil13}. \seb{This expression is particularly convenient because the function} $\zeta(\nu)$ can be \seb{expressed as a function of the asymptotic properties of pure p and g modes} as \citep{Mosser2015, Gehan2018}
\begin{equation}
    \zeta = \left[1+\frac{\nu^2}{q}\frac{\Delta\Pi_1}{\Delta\nu}\frac{1}{\frac{1}{q^2}\sin^2\theta_\mathrm{p}+\cos^2\theta_\mathrm{p}}\right]^{-1}. \label{eq:zeta_function}
\end{equation}

Thus, the frequencies of $m=0$ modes $\nu_{m=0}$ can be first obtained by solving Eq.~\ref{eq:asymptotic_expression}, and the frequencies of $m=\pm1$ modes can then be calculated as $\nu_m=\nu_0 - m\delta\nu_\mathrm{rot}$. The advantages of this approach are that the best-fitting result provides the splitting identification even though the splittings overlap each other, and the computation is relatively fast since the $\zeta$ function is explicit. 

\LG{The disadvantage of Approach One is the lack of consideration of near-degeneracy effects (NDE). NDE arise when the frequency separation between consecutive mixed modes is comparable to or lower than the rotational splitting. They generally lead to asymmetries in the rotational multiplets of p-dominated mixed modes (\citealt{Deheuvels2017}). As stars evolve along the RGB, we expect NDE to become more important because their asymptotic period spacing decreases. Also, NDE are naturally larger for stars with faster-rotating cores. Approach One uses the $\zeta$ function to obtain approximate expressions for the rotational splittings (Eq.~\ref{eq:asymptotic_expression}) and therefore has symmetric rotational multiplets by construction. For stars that show non-negligible NDE, Approach One inadequately captures the structure of p-dominated modes, which can lead to biases in the measurements of the envelope rotation $\Omega_\mathrm{env}$ and the coupling factor $q$. }

\subsubsection{Approach Two}\label{subsec:approach2}

\LG{Approach Two consists of including rotational perturbations to the asymptotic frequencies of pure p- and g-modes before accounting for the coupling between the cavities. These frequencies are thus expressed as}
\begin{equation}
    \nu_{\mathrm{g}, m} = \nu_\mathrm{g} - m \frac{\Omega_\mathrm{core}}{4\pi},
\end{equation}
and
\begin{equation}
    \nu_{\mathrm{p}, m} = \nu_{\mathrm{p}, l=1} - m \frac{\Omega_\mathrm{env}}{2\pi}.
\end{equation}
They are then used to calculate the perturbed g and p phase terms for $m=1,0, -1$ as
\begin{equation}
    \theta_{\mathrm{g}, m} = \pi \frac{1}{\Delta\Pi_1}\left(\frac{1}{\nu}-\frac{1}{\nu_{\mathrm{g}, m}}\right),
\end{equation}
and
\begin{equation}
    \theta_{\mathrm{p}, m}=\pi\frac{\nu-\nu_{\mathrm{p}, m}}{\Delta\nu}.
\end{equation}
After this, the implicit function for $m=1$, $0$, and $-1$ is solved three times to obtain the frequencies with different $m$ values:
\begin{equation}
    \tan \theta_{\mathrm{p}, m } = q \tan \theta_{\mathrm{g}, m},~~\mathrm{for}~~m=1,~0,~-1. \label{eq:tan_tan_three_times}
\end{equation}
Solving eq.~\ref{eq:tan_tan_three_times} is much more time-consuming than Approach One since it is implicit, but we argue below that it provides a more reliable solution. We note that Approach Two has already been used by \cite{Li2022Nature}, \cite{Deheuvels2023}, and \cite{Li2023}, but these works also included magnetism-induced perturbations. Here we only consider rotational perturbations.

\seb{Approach Two is in fact similar to the one proposed by \cite{ong22} to account for the NDE. Starting from the oscillation operator perturbed by rotation, the authors isolate pure pressure modes ($\pi$ modes) and pure gravity modes ($\gamma$ modes). Then, they decompose the mode eigenfunctions in the basis of isolated $\pi$ and $\gamma$ modes and they argue that the rotation operator is diagonal in this representation. This shows that in this basis, the effects of rotation on the mode frequencies can indeed be separated from the effects of the coupling, as we have done in Approach Two. \cite{ong22} showed that they obtained the same expressions as the ones that were found by \cite{Deheuvels2017} to account for the NDE. The only difference between the approach of \cite{ong22} and our Approach Two is that the latter uses WKB expressions of mode frequencies, while the former uses eigenfrequencies and eigenfunctions numerically computed with a code solving the unperturbed set of oscillation equations. We thus expect Approach Two to adequately account for the NDE, contrary to Approach One.}

%\LG{Approach Two has been used by \cite{Li2022Nature} and \cite{Li2023}, but these works also included magnetism-induced perturbations. Here we only consider rotational perturations.} Instead of using the $\zeta$ function, we added rotational perturbations on the pure g and p modes, as
%\begin{equation}
%    \nu_{\mathrm{g}, m} = \nu_\mathrm{g}\mp m \frac{\Omega_\mathrm{g}}{4\pi},
%\end{equation}
%and
%\begin{equation}
%    \nu_{\mathrm{p}, m} = \nu_{\mathrm{p}, l=1}\mp m \frac{\Omega_\mathrm{p}}{2\pi},
%\end{equation}
%then calculated the perturbed g and p phases for $m=1,0, -1$ as
%\begin{equation}
%    \theta_{\mathrm{g}, m} = \pi \frac{1}{\Delta\Pi_1}\left(\frac{1}{\nu}-\frac{1}{\nu_{\mathrm{g}, m}}\right),
%\end{equation}
%and
%\begin{equation}
%    \theta_{\mathrm{p}, m}=\pi\frac{\nu-\nu_{\mathrm{p}, m}}{\Delta\nu}.
%\end{equation}
%After that, we solved the implicit function for $m=1$, $0$, and $-1$ three times to obtain the frequencies with different $m$ value:
%\begin{equation}
%    \tan \theta_{\mathrm{p}, m } = q \tan \theta_{\mathrm{g}, m}~~\mathrm{for}~~m=1,~0,~-1. \label{eq:tan_tan_three_times}
%\end{equation}

%Solving eq.~\ref{eq:tan_tan_three_times} is much more time-consuming since it is implicit, while it provides a more reliable solution compared to the Approach one (discussed below) since some approximations were used when deducing the $\zeta$ function (Eq.~\ref{eq:zeta_function}).

\subsection{MCMC optimisation for mixed modes} \label{sec:MCMC_mixed_mode}

\begin{table}[]
\small
        \caption{Ten parameters and their ranges used in the MCMC fitting described in Sect.~\ref{sec:MCMC_mixed_mode}. }
    \label{tab:six_parameters}
\centering
    \begin{tabular}{ll}
    \hline
    Parameter name & range \\
    \hline
    Large separation frequency $\Delta \nu$ & no constraint \\
    Pressure mode phase $\varepsilon_\mathrm{p}$ & no constraint \\
    Second term coefficient $\alpha$ & no constraint \\
    Small separation parameter $D$ & no constraint \\
    Asymptotic period spacing $\Delta\Pi_1$ & $[\Delta \Pi_\mathrm{1,init}-0.5\,\mathrm{s}, \Delta \Pi_\mathrm{1,init}+0.5\,\mathrm{s}]$ \\
    Coupling factor $q$ & [0.08, 0.25] \\
    Phase of g mode $\varepsilon_g$ & [-0.1, 1.1] \\
    Dipole ($l=1$) frequency shift $f_\mathrm{shift}$ & [0.3$\,\mathrm{\mu Hz}$, 1.0$\,\mathrm{\mu Hz}$] \\
    Core angular frequency $\Omega_\mathrm{core}$ & [0$\,\mathrm{\mu Hz}$, 20$\,\mathrm{\mu Hz}$] \\
    Envelope angular frequency $\Omega_\mathrm{env}$ & [-1$\,\mathrm{\mu Hz}$, 1$\,\mathrm{\mu Hz}$] \\
    \hline
    \end{tabular}
     \tablefoot{$\Delta \Pi_\mathrm{1,init}$ is the initial guess of the period spacing from the stretched \'{e}chelle diagram.}
\end{table}

We ran an
%similar 
MCMC optimisation algorithm to search for the best-fitting parameters of $l=0$, $l=2$, and $l=1$ frequencies. Table~\ref{tab:six_parameters} lists all the ten parameters that were optimised, along with their ranges of uniform priors. The top four parameters describe the p-mode asymptotic expression in Eq.~\ref{eq:p_mode_asymptotic_relation}, while the rest six parameters describe the mixed modes. The likelihood function for $l=1$ mixed modes is defined as:
\begin{equation}
    \ln L_{l=1} = -\frac{1}{2}\sum_{m=1, 0, -1} \sum_i\left[\frac{\left(\nu_{m, i}^\mathrm{obs}-\nu_{m, i}^\mathrm{cal}\right)^2}{\sigma_{m, i}^2} +\ln \left(2\pi\sigma_{m, i}^2\right)\right], \label{eq:likelihood}
\end{equation}
where $\nu_{m, i}^\mathrm{obs}$ and $\nu_{m, i}^\mathrm{cal}$ are the observed and calculated $i^\mathrm{th}$ frequencies with azimuthal order $m$, and $\sigma_{m, i}$ is the uncertainty of the $i^\mathrm{th}$ frequency with azimuthal order $m$. 

\seb{We followed a two-step procedure.} In the first step, we ran the MCMC optimisation code with fixed p-mode parameters and constant uncertainties $\sigma_{m,i}^\mathrm{obs}=0.01\,\mathrm{\mu Hz}$ for $l=1$ mixed mode frequencies. \seb{At this point,} we \seb{considered} the points with a signal-to-noise ratio (S/N) larger than ten as input frequencies. The best-fitting results helped us identify the rotational splittings of $l=1$ mixed modes. We then fitted Lorentzian profiles on them, which allowed us to derive \seb{proper estimates of the mode frequencies and their related} uncertainties $\sigma_{m, i}^\mathrm{obs}$. In Appendix \ref{app_spectrum_8636389}, fig.~\ref{fig:whole_spectrum_fit_of_8636389} \seb{shows an example of the result of the Lorentzian fit for a star of the sample.}
%displays the result of the splitting fitting. 
%We find that our Lorentzian fit reproduced the observed spectrum well. 
%The oscillation frequencies and their uncertainties were thus obtained and will be used in the following fitting. 
%In the Lorentzian fitting of the splittings, 
\seb{Beside the central frequency, this fitting procedure also provides estimates of} the rotational splittings $\delta \nu_\mathrm{rot}$, the stellar inclination $i$, the mode amplitudes $a_0$ and linewidths $\eta$, and the background of the power spectrum. 

In the second step, we re-ran the MCMC code using the frequencies and uncertainties acquired through the Lorentzian fit. This step enabled us to accurately estimate the uncertainties \seb{of the parameters related to the asymptotic expressions of the modes}. We then treated the p-mode parameters as free parameters, incorporating them into the MCMC process alongside the mixed-mode parameters. 
We found that our best-fitting models have a residual spread, whose median is larger when $|\theta_\mathrm{p}|< 0.15\pi$ (corresponding to p-dominated modes) and smaller when $|\theta_\mathrm{p}|> 0.15\pi$ (which are g-dominated modes). \seb{We indeed found a median $\sigma_{\rm p} = 14$~nHz for p-dominated modes, and $\sigma_{\rm g} = 7$~nHz for g-dominated modes.} We consider this residual spread as the model uncertainty of the asymptotic expressions. The larger residual on p-dominated modes might be a hint of small glitches, which influence our estimation of envelope rotation rates. To include the model uncertainty properly, we define the frequency uncertainty as 
\begin{equation}
\begin{split}
    & \sigma_{m, i} = \left[ \sigma_{\rm p}^2+ \left(\sigma_{m, i}^\mathrm{obs}\right)^2 \right]^{0.5}~\mathrm{when}~|\theta_\mathrm{p}|<0.15\pi, \\
    & \mathrm{or} \\
    & \sigma_{m, i} = \left[ \sigma_{\rm g}^2+ \left(\sigma_{m, i}^\mathrm{obs}\right)^2 \right]^{0.5}~\mathrm{when}~|\theta_\mathrm{p}|>0.15\pi,
\end{split}
\end{equation}
where $\sigma_{m, i}^\mathrm{obs}$ is the observed frequency uncertainty from the Lorentzian profile fitting. To fit p modes and mixed modes together, the likelihood function is defined as the sum of Eqs.~\ref{eq:likelihood} and \ref{equ:likelihood_for_p_mode}, that is
\begin{equation}
    \ln L = \ln L_\mathrm{p} + \ln L_{l=1}.
\end{equation}

The MCMC code for mixed mode is more time-consuming. We used 22 parallel chains and 6000 steps, and we discarded the first half of the chain to obtain the posterior distributions. This choice ensures both MCMC convergence and time cost. We also visually inspected all the results of stars to ensure that the chains were well converged and their posterior distributions did not reach the boundary of their uniform priors. In some cases, we found that $\varepsilon_\mathrm{g}$ reached its boundary (the initial one is $-0.1 < \varepsilon < 1.1$), so we just set a larger boundary for the uniform prior of $\varepsilon_\mathrm{g}$. 

\begin{figure*}
    \centering
    \includegraphics[width=1\linewidth]{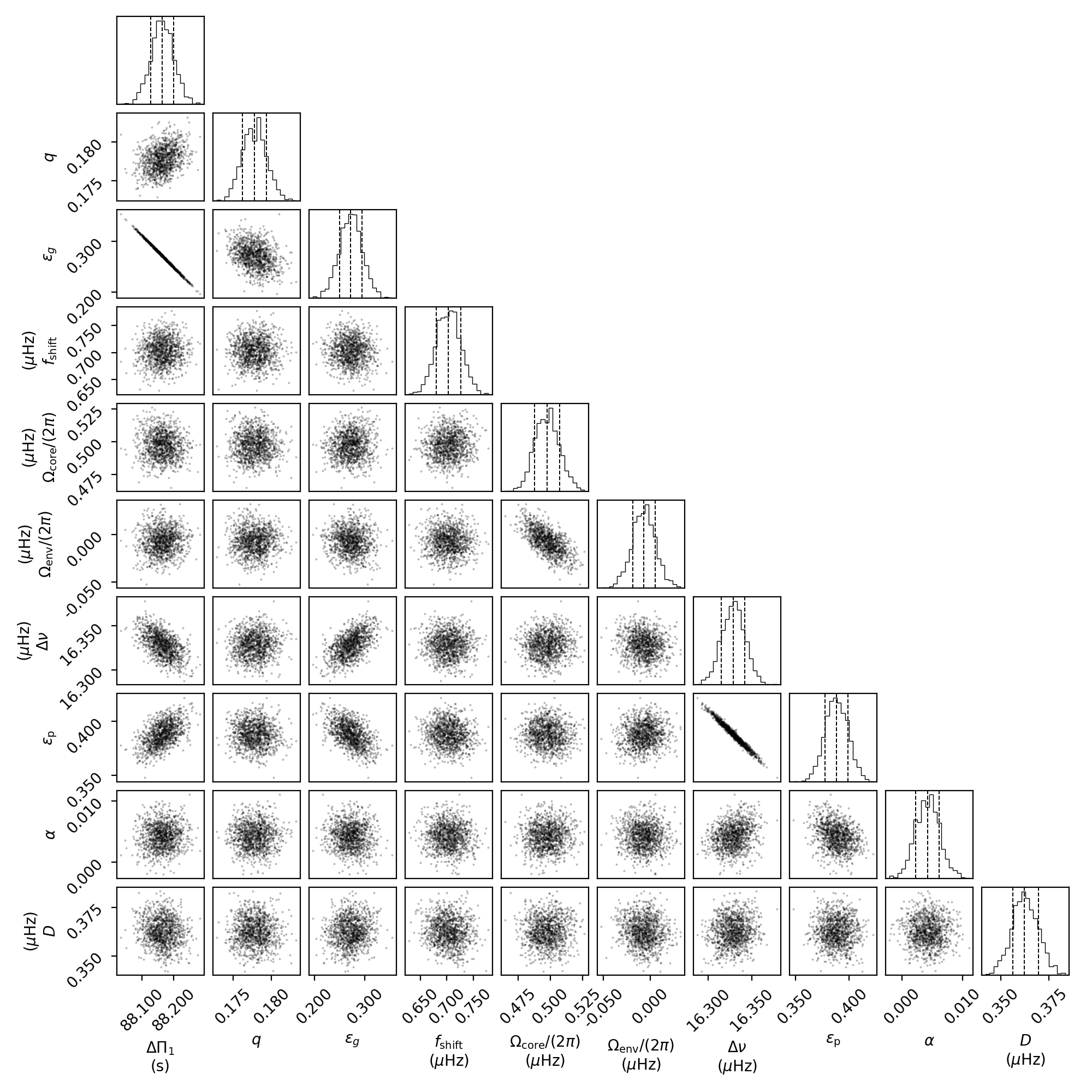}
    \caption{Corner diagram of the posterior distributions of KIC\,9289599 by Approach Two.}
    \label{fig:corner_KIC_9289599}
\end{figure*}

Here we show KIC\,9289599 as an example, which is selected randomly from our sample. Figure~\ref{fig:corner_KIC_9289599} displays the corner diagram of the MCMC fit using Approach Two, which demonstrates that the chain converges well. 
\LG{We find that there is a strong anti-correlations between $\Delta\Pi_1$ and $\varepsilon_\mathrm{g}$, which can be understood from Eq.~\ref{eq:g_mode_equally_spacing_formula} as the product $\Delta \Pi_1 \varepsilon_\mathrm{g}$ plays a holistic role. The core and envelope rotation rates ($\Omega_\mathrm{core}$ and $\Omega_\mathrm{env}$) also show anti-correlations. This is because the sum of the two rates, weighted by the $\zeta$ function, determines the magnitude of the rotational splitting. Some correlations between p-mode parameters and g-mode parameters are also found, such as the correlation between $\Delta \Pi_1$ and $\Delta \nu$, which can be understood in Eq.~\ref{eq:asymptotic_expression}. }

\subsection{Sample and selection effects}

\begin{figure}
    \centering
    \includegraphics[width=1\linewidth]{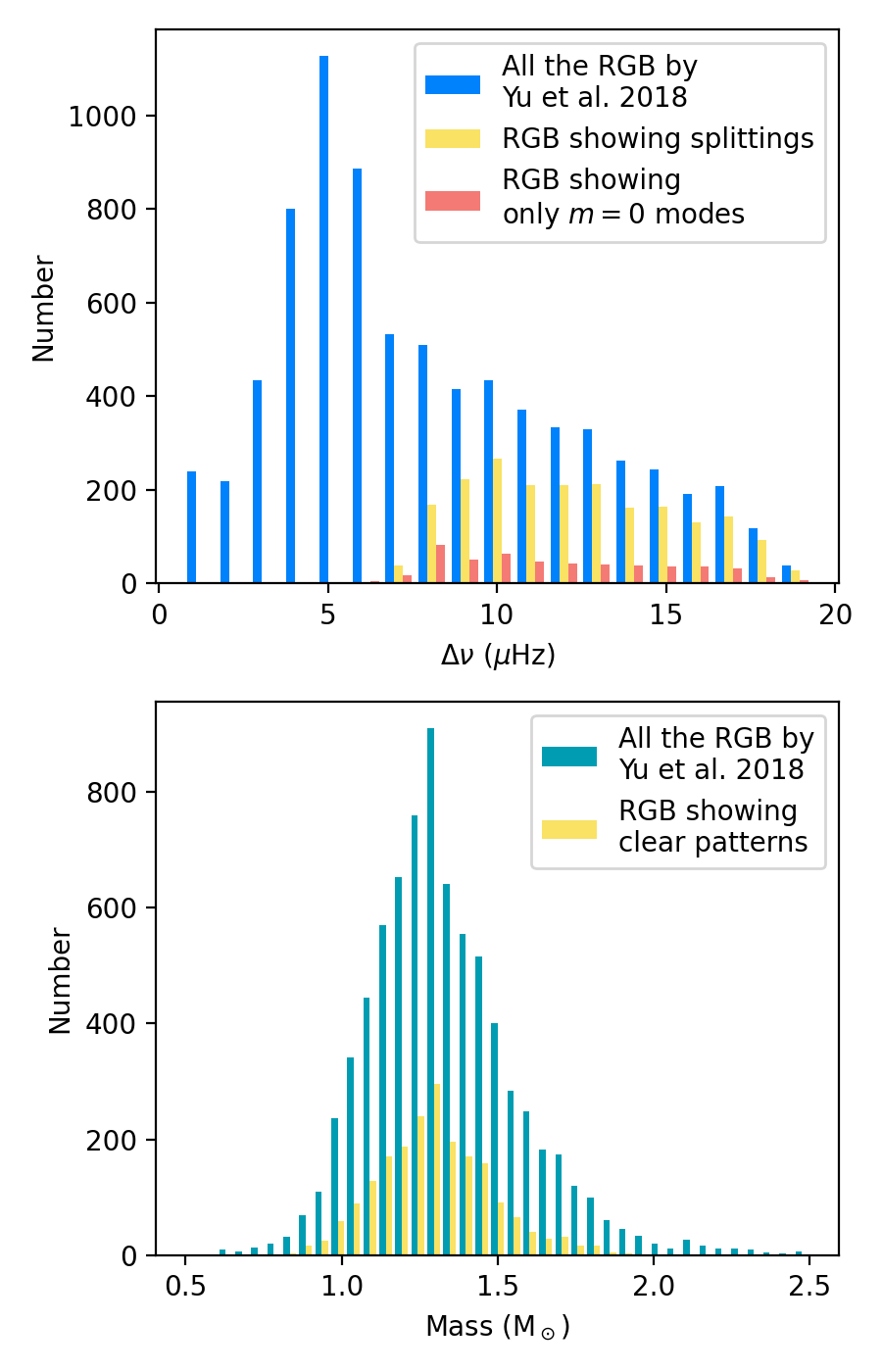}
    \caption{Distributions of $\Delta\nu$ (top) and mass (bottom) of our sample and the sample by \cite{Yu2018}. }
    \label{fig:sample_distributions}
\end{figure}

During the visual inspection of our stars, we removed the stars that show glitches caused by the discontinuity in the cores \citep{Cunha2015}, and also the stars showing curved stretched \'{e}chelle diagrams or asymmetric rotational splittings which are caused by their central magnetic fields \citep{Li2022Nature, Deheuvels2023, Li2023}. 
In total, we collected \totalstarnumber~stars that show clear $l=1$ mixed-mode patterns. \rotationnumber~stars show rotational splittings hence the ten parameters (Table~\ref{tab:six_parameters}) including the core and the envelope rotation rates are measured. Among them, \doubletnumber~stars show doublets and \tripletnumber~stars show triplets. However, \norotationnumber~stars show only $l=1,~m=0$ modes. In these cases, we could not measure their internal rotations, so we only measured the \jb{eight other parameters.}  %jb %rest eight parameters except core and envelope rotations. 
\LG{In solar-like oscillations, the relative amplitudes of the components in rotational splittings are determined by the inclination of the rotational axis \cite{Gizon2003}. Inclination close to $0^\circ$ leads to the visibility of $m=0$ modes while a $90^\circ$ inclination allows $m=\pm1$ modes to be visible. \seb{Intermediate} inclinations generate triplets. We also measured the inclinations of our sample, which are discussed in Appendix~\ref{appendix_sec:other_parameters}. }
All the \totalstarnumber~stars were inspected to assure that the best-fitting results are correct. \LG{During the inspection, we concentrated on evaluating the following points: 1) whether the fit residuals are small; and 2) whether the MCMC algorithm converged properly. The main issues with the above two points are primarily related to mode identifications, such as failing to correctly identify the large frequency separation, mistaking $l=3$ modes for $l=1$ mixed modes, or encountering issues while fitting the Lorentzian profile for rotational splitting. These problems could all be manually corrected. }

Figure~\ref{fig:sample_distributions} displays the distributions of $\Delta \nu$ and stellar mass of our sample and the sample by \cite{Yu2018}. For the $\Delta\nu$ distribution in the top panel of Fig.~\ref{fig:sample_distributions}, we observe a distinct cut-off at around $\Delta \nu \approx7.5\,\mathrm{\mu Hz}$. 
\LG{Below this value, the mixed modes are unclear due to the short mode lifetime caused by the radiative damping \citep{Grosjean2014}, hence it is hard to identify any pattern. We find that our $\Delta \nu$ cut-off of \LG{$\sim 7.5\,\mathrm{\mu Hz}$} is \LG{larger} than what is predicted by \citep{Grosjean2014}, which is $4.9\,\mathrm{\mu Hz}$.}
The bottom panel of Fig.~\ref{fig:sample_distributions} displays the distribution of stellar mass in our sample and the sample by \cite{Yu2018}. The masses of the stars in our sample predominantly range from approximately $\sim0.7$ to $2.0$\,$\mathrm{M_\odot}$. The fraction of stars exhibiting clear mixed-mode pattern increases with masses, and reaches the maximum around 1.25\,$\mathrm{M_\odot}$, where about 30\% stars are included in our sample. Subsequently, the fraction gradually decreases, reaching around 5\% at $M\approx2.0$\,$\mathrm{M_\odot}$. \LG{The scarcity of high-mass stars could be attributed to the suppression of $l=1$ modes induced by central magnetic fields, whose prevalence increases with increasing stellar mass \citep{Stello2016}.}

\section{Result comparison}\label{sec:result_comparison}

\subsection{
%Approach comparison and the NDE
Approach One vs Approach Two}\label{subsec:Approach_comparison}

\begin{figure*}
    \centering
    \includegraphics[width=1\linewidth]{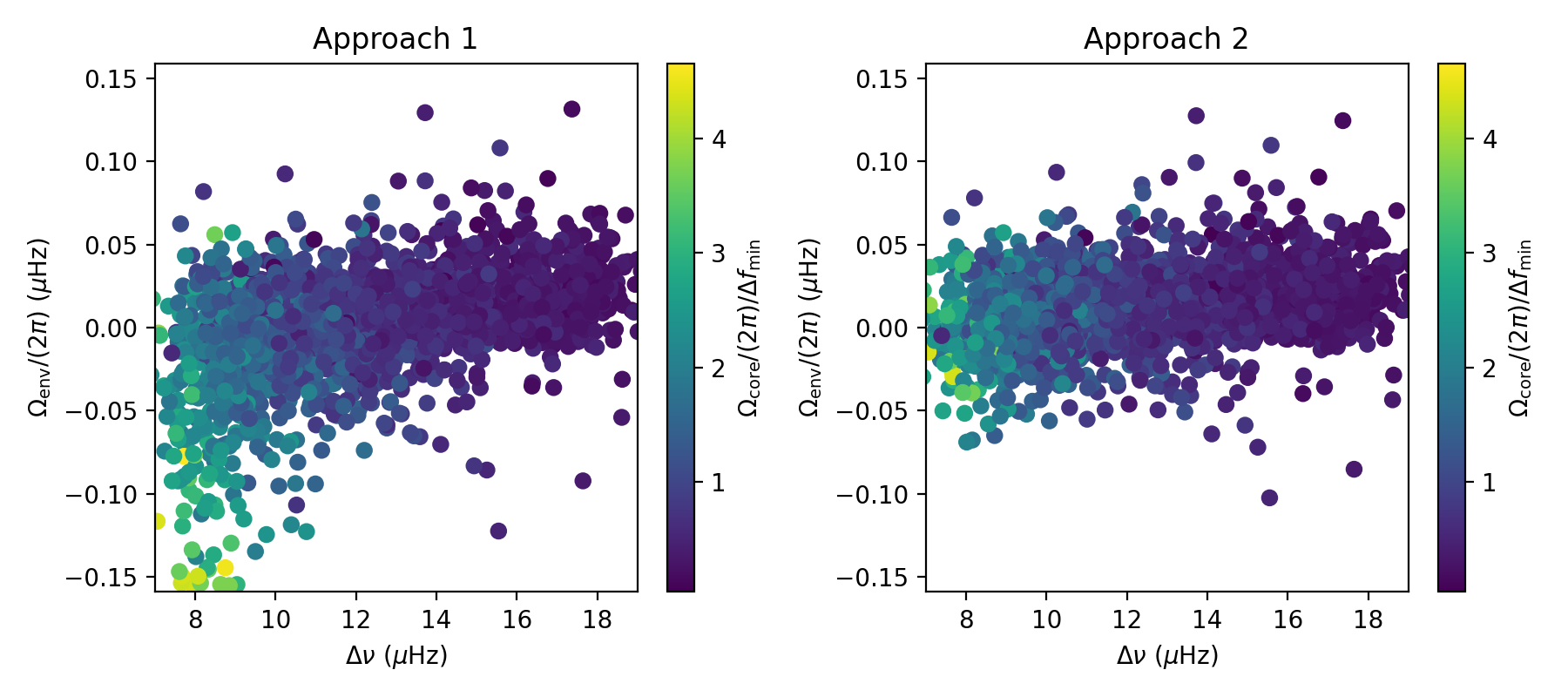}
    \caption{Envelope rotation rates derived by Approaches One and Two. %\LG{$\Delta f_\mathrm{min}$ represents the smallest frequency separation of $l=1$ mixed modes (see Sect.~\ref{subsec:Approach_comparison}). 
    The colour bar
    \seb{indicates the intensity of NDE, calculated as the ratio between the core rotation and $\Delta f_\mathrm{min}$, which is the smallest frequency separation between consecutive $l=1$ mixed modes around $\nu_\mathrm{max}$ (see Sect.~\ref{subsec:Approach_comparison}.}
    %shows the ratio between the core rotation and $\Delta f_\mathrm{min}$, therefore representing the extent of the NDE.} We note that Approach One reports some negative envelope rotations when $\Delta\nu \lesssim 11\,\mathrm{\mu Hz}$.  
    }
    \label{fig:envelope_rotation_by_different_approach}
\end{figure*}

\begin{figure*}
    \centering
    \includegraphics[width=1\linewidth]{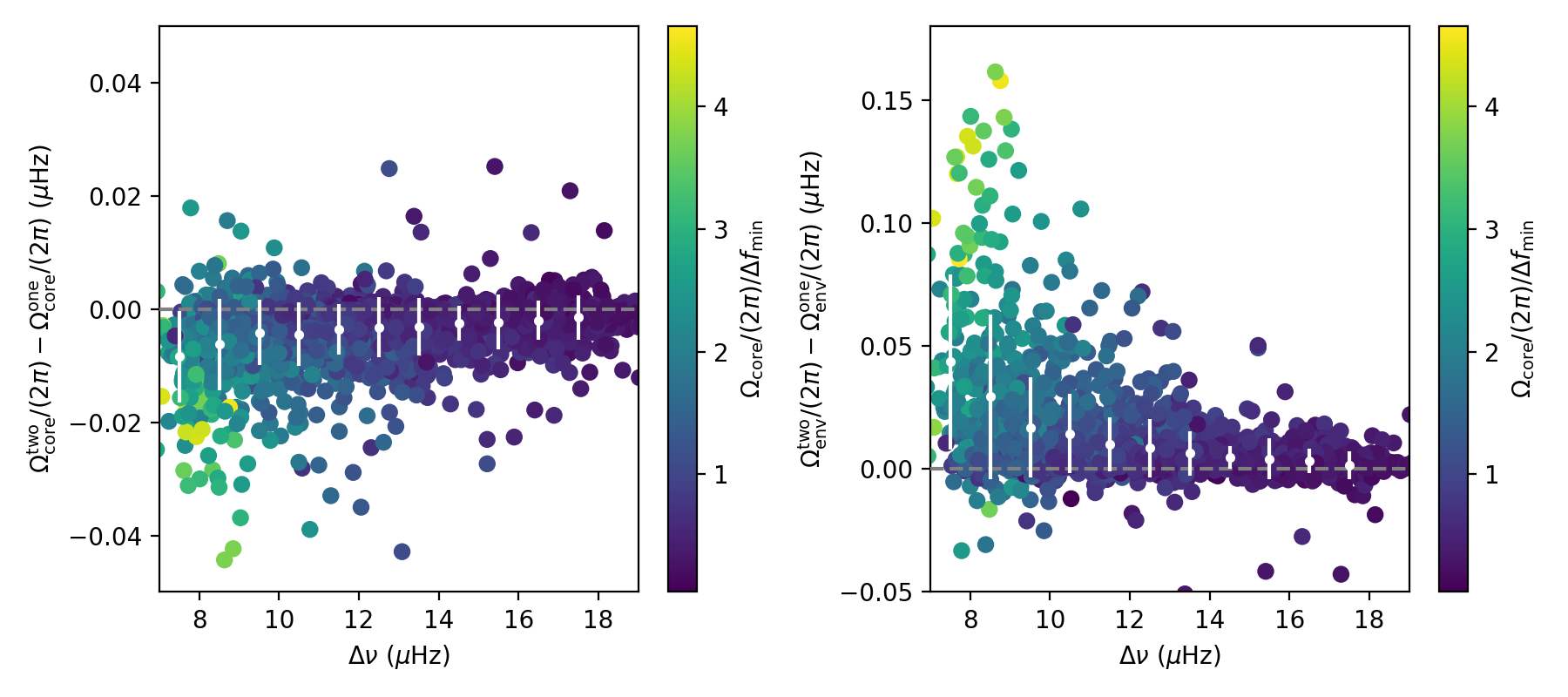}
    \caption{The differences of the core and envelope rotation rates between Approach Two and One (defined as the results from Approach Two minus the results from Approach One). \LG{The colour bar represents the extent of the NDE (see Fig.~\ref{fig:envelope_rotation_by_different_approach} and Sect.~\ref{subsec:Approach_comparison}).} The horizontal dashed lines mark the location where the values are equal. The white error bars show the mean and standard deviations in each bin of $\Delta \nu$. }
    \label{fig:difference_between_approach_two_and_one}
\end{figure*}

We compared the results given by the two different approaches \seb{used to compute the rotational splitting of mixed modes} in Sect.~\ref{subsec:approach1} and \ref{subsec:approach2}. %, and \ref{subsec:approach3}. T
%The other parameters show general consistency between different approaches, while the discrepancy emerges from the measurements of their envelope rotation rates $\Omega_\mathrm{env}$. 
\seb{These two approaches show general consistency for the estimates of all parameters but one, the average envelope rotation rate $\Omega_\mathrm{env}$.}
As shown in the two panels of Fig.~\ref{fig:envelope_rotation_by_different_approach}, Approach One (left panel) tends to yield \seb{significantly} negative rotation rates of the envelopes.
\seb{This is} particularly \seb{the case} when $\Delta \nu<11\,\mathrm{\mu Hz}$\seb{, that is, for more evolved giants}. In contrast, Approach Two (right panel of Fig.~\ref{fig:envelope_rotation_by_different_approach}) does not exhibit any spread of $\Omega_\mathrm{env}$ \seb{for stars with}  small $\Delta\nu$. 
%We therefore claim that the envelope rotation rates measured with the $\zeta$ function have systematic deviations. Consequently, we recommend approach two as the reliable method for measuring envelope rotations, and we have used the results obtained by approach two in this study.

In Fig.~\ref{fig:difference_between_approach_two_and_one}, we show the differences %\LG{of} core and envelope rotation rates obtained using Approach One and Approach Two. 
\jb{obtained between Approach One and Two for core and envelope rotation rates.}
The left panel of Fig.~\ref{fig:difference_between_approach_two_and_one} shows that there is a slight \seb{but significant} systematic deviation in the core rotation rates, particularly at smaller values of $\Delta \nu$. At $\Delta\nu\approx8\,\mathrm{\mu Hz}$, the core rotation rates given by Approach Two are \seb{on average smaller than those given by Approach One by about 10~nHz.}
%$-0.01\,\mathrm{\mu Hz}$ . 
However, this difference diminishes as $\Delta \nu$ increases. Regarding the envelope rotation rates (right panel in Fig.~\ref{fig:difference_between_approach_two_and_one}), 
%apart from the extremely negative rotation rates given by approach one, 
the two approaches %still 
show a systematic deviation, \seb{accentuated at small values of $\Delta\nu$ by the subset of stars that are found to have large negative rotation rates with Approach One. }
%Approach two tends to report slightly higher values for the envelope rotations, which are considered to be more physically reliable. Consequently, we confirm that approach two provides better results, particularly in terms of envelope rotation measurements.

We suspected that these differences 
%(particularly for the envelope rotation measurements) 
were caused by near-degeneracy effects (NDE). \seb{To verify this, we estimated the intensity of the NDE that is expected in the targets of our sample. For this purpose, we calculated the smallest frequency separation $\Delta f_\mathrm{min}$ between consecutive mixed modes around $\nu_\mathrm{max}$ and we compared this quantity to the core rotation rate $\Omega_\mathrm{core}/(2\pi)$. NDE become important when the ratio $\Omega_\mathrm{core}/(2\pi\Delta f_\mathrm{min})$ becomes of the order of unity or more. In Figs.~\ref{fig:envelope_rotation_by_different_approach} and \ref{fig:difference_between_approach_two_and_one}, the points are colour-coded with the value of this ratio. It appears clearly that the stars that are found to have large negative envelope rotation rates with Approach One correspond to stars for which large NDE are expected. Conversely, when using Approach Two, stars with large NDE have envelope rotation rates that are similar to those of other stars (right panel of Fig.~\ref{fig:difference_between_approach_two_and_one}).}
%e extent of the NDE in the stars, we calculated the smallest frequency separation ($\Delta f_\mathrm{min}$) from the best-fitting asymptotic frequencies, and show the ratio $\Omega_\mathrm{core}/\Delta f_\mathrm{min}$ in Figs.~\ref{fig:envelope_rotation_by_different_approach} and \ref{fig:envelope_rotation_by_different_approach}. We find that the stars with a large ratio $\Omega_\mathrm{core}/\Delta f_\mathrm{min}$ indeed exhibit negative envelope rotation rates, showing that the lack of consideration of the NDE in Approach One indeed leads to a deviation of surface rotation measurements. 

\begin{figure}
    \centering
    \includegraphics[width=\linewidth]{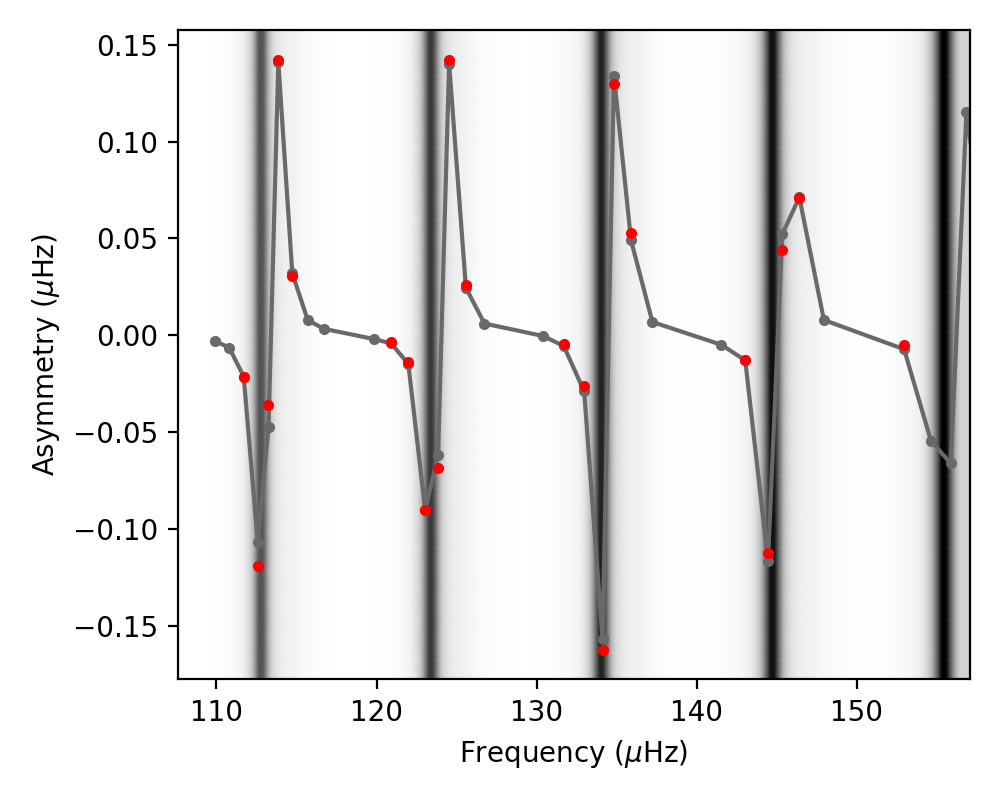}
    \caption{Synthetic asymmetries in $l=1$ rotational multiplets resulting from near-degeneracy effects in KIC\,11245496. The red dots are the simulated asymmetries following the method by \cite{Deheuvels2017}, while the grey dots and lines are the best-fitting results by Approach Two. The background shows the $\zeta$ values: the darker, the smaller, hence the dark fringes stand for the positions of p-dominated modes. }
    \label{fig_asym_NDE}
\end{figure}

\seb{As an additional verification that Approach Two adequately captures NDE, we considered a reference star for which NDE are expected to be non-negligible and Approach One reports a large negative envelope rotation rate (we picked KIC\,11245496, for which $\Omega_\mathrm{core}/(2\pi\Delta f_\mathrm{min}) = 2.4$). We computed a stellar model approximately representative of this star (showing similar values of $\Delta\nu$ and $\Delta\Pi_1$) with the MESA code (\citealt{paxton11}) and we calculated its mode frequencies and eigenfunctions with \textsc{ADIPLS} (\citealt{adipls}). To estimate the rotational perturbation to the mode frequencies accounting for NDE in this model, we followed \cite{Deheuvels2017} \LGvtwo{to generate synthetic frequencies} (see Appendix \ref{app_NDE}). Significant multiplet asymmetries arise in the vicinity of p-dominated modes owing to NDE, as can be seen in Fig.~\ref{fig_asym_NDE}. We then considered the perturbed frequencies of this model as mock observations and attempted to fit them using Approach Two. The results are shown in Fig.~\ref{fig_asym_NDE}, where it can be seen that the asymmetries of multiplets are very well reproduced using Approach Two (for comparison, we recall that the asymmetries are by construction equal to zero when using Approach One). Also, the parameters recovered when using Approach Two (in particular the core and envelope rotation rates) are in good agreement with the input values. This confirms that approach indeed accounts well for NDE. }\LGvtwo{In Fig.~\ref{fig_asym_NDE_real} of appendix \ref{appendix_sec:read_asymmetries_in_11245496}, we present the observed splitting asymmetries and the best-fitting results of KIC\,11245496. We claim that Approach Two indeed can reproduce the observed splitting asymmetries. }

%\begin{figure*}
%    \centering
%    \includegraphics[width=0.9\linewidth]{figures/near_degeneracy_effect_comparison.png}
%    \caption{Result comparisons for envelope rotations in different stellar models and different codes by Li or Ballot. \LG{remove Ballot result} }
%    \label{fig:result_comparison_Li_Ballot}
%\end{figure*}

%\begin{table}[]
%    \centering
%    \caption{Pamareters of the six models that were used to test the effectiveness of different approaches.}\label{tab:simulated_model}
%    \begin{tabular}{llll}
%    \hline
%    Case number & $\Omega_\mathrm{core}/(2\pi)$ & $\Omega_\mathrm{env}/(2\pi)$ & Comment\\
%     & ($\mu$Hz) & ($\mu$Hz) & \\
%    \hline
%    000     &  0.479 & -0.06 & negligible NDE \\
%    001     &  0.479 & 0.03 & negligible NDE \\
%    002     &  0.479 & 0.143 & negligible NDE \\
%    003 & 1428 & -0.06 & large NDE \\
%    004 & 1428 & 0.03 & large NDE \\
%    005 & 1428 & 0.143 & large NDE\\
%    \hline
%    \end{tabular}

%\end{table}

%Description of the results from Approach One and Approach Two applied to this case.

\subsection{Comparison with previous works}\label{subsec:comparison}

\begin{figure*}
    \centering
    \includegraphics[width=0.9\linewidth]{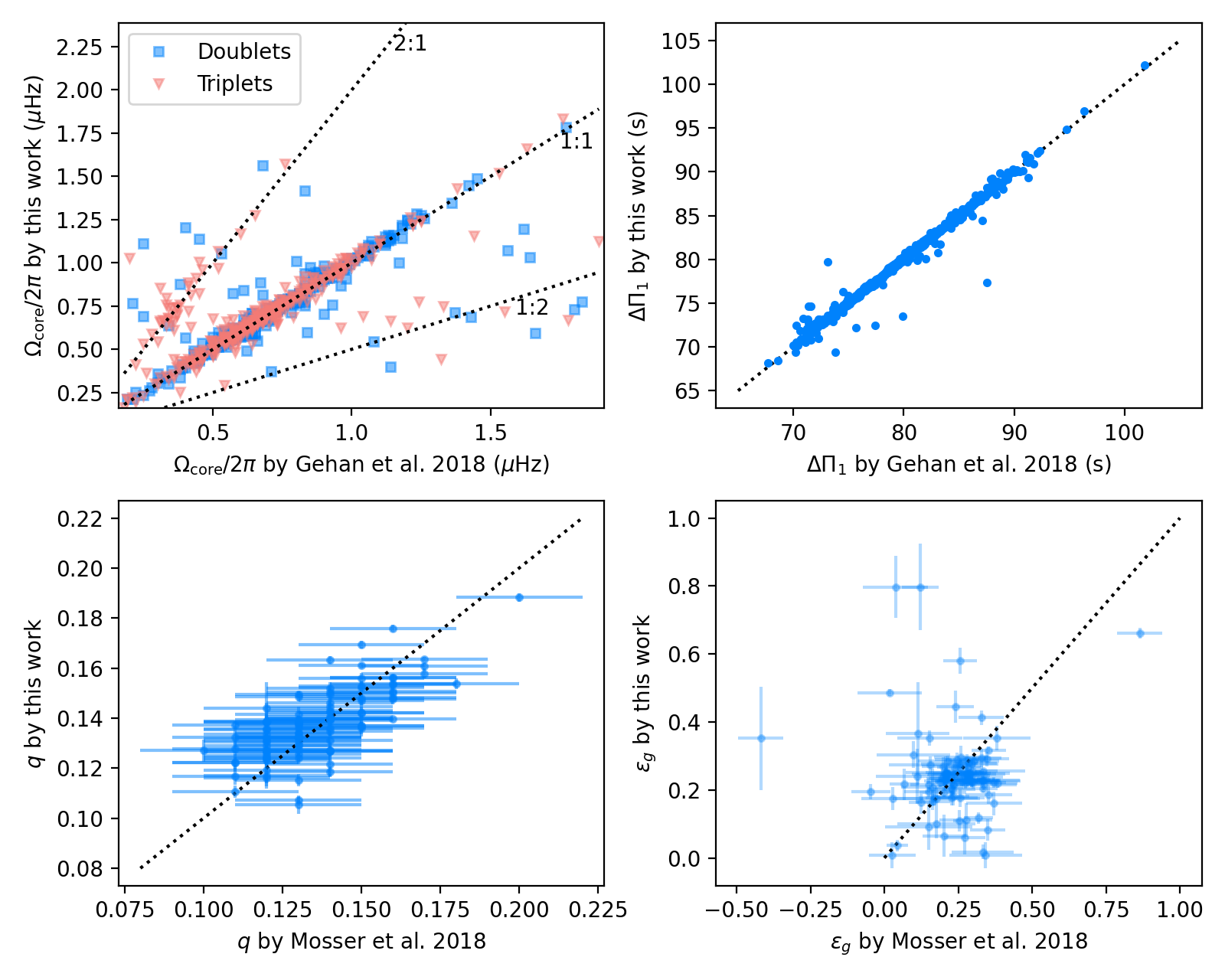}
    \caption{Comparison between our result with previous studies. Top left: core rotation rate $\Omega_\mathrm{core}$. Top right: period spacing $\Delta\Pi_1$. Bottom left: coupling factor $q$. Bottom right: g-mode phase $\varepsilon_\mathrm{g}$. the dashed line in each panel shows the 1:1 relation, or 2:1 and 1:2 as marked. }
    \label{fig:result_comparison}
\end{figure*}

We selected the stars that were also studied by \cite{Gehan2018} and \cite{mosser18} and compared the parameters obtained in our work with those from the previous studies. In Fig.~\ref{fig:result_comparison}, we present the comparison between the results obtained using approach two and the previous studies.

In the top left panel of Figure~\ref{fig:result_comparison}, we present the comparison of the core rotation rates ($\Omega_\mathrm{core}$) between our results and those from \cite{Gehan2018}. For this comparison, we selected a total of 693 stars that are studied in both works. In general, the measurements of core rotation rates show good agreement between the two studies, indicating consistency. However, we do observe a small fraction of stars where the core rotation rates obtained in our study are approximately twice those reported by \cite{Gehan2018}. \LGnew{To be specific, there are 32 stars that \cite{Gehan2018} reported larger core rotation rates ($\Delta \Omega_\mathrm{core}/2\pi$ larger than 0.1\,$\mu \mathrm{Hz}$), while 61 stars were reported smaller core rotation rates by \cite{Gehan2018} ($\Delta \Omega_\mathrm{core}/2\pi$ smaller than 0.1\,$\mu \mathrm{Hz}$). }
This discrepancy could be attributed to the misidentification of the azimuthal order $m$ in these particular stars. As plotted by the grey downward-pointing triangles in the upper left panel of Figure~\ref{fig:result_comparison}, a significant portion of stars that fall along the 2:1 trend exhibit triplets in their rotational splittings. It is plausible that the previous study misidentified these triplets as doublets, failing to recognise one of the components, consequently leading to the calculation of core rotation rates at half their actual values. Additionally, we find several stars in our sample that exhibit smaller rotation rates compared to the measurements reported in \cite{Gehan2018}. The stars located along the 1:2 trend might be a result of the previous work misidentifying doublets ($m=\pm1$) as parts of triplets (comprising components with $m=1$ and $0$, or $m=0$ and $-1$). %These differences may arise from various factors such as improved data analysis techniques or different observational characteristics taken into account during the analysis.

In the top right panel of Fig.~\ref{fig:result_comparison}, we present the comparison of the g-mode period spacing ($\Delta\Pi_1$) between our results and those reported in \cite{Gehan2018}. For this comparison, we selected a total of 765 stars that are common to both studies. Our measurements of the g-mode period spacing show good consistency with the previous study, as the majority of stars exhibit similar values in both studies. Only six stars demonstrate significant differences in their period spacings between the two studies. \LG{Among these six stars, KIC\,1865102 exhibits significant scatters that deviate from the best-fitting asymptotic frequencies. KIC\,4667909, 5255835, 12207840, 6579495, and 8418309 display high core rotation rates, indicating that the multiple ridges in the stretched \'{e}chelle diagram have significantly different period spacings, with some even crossing over to another ridge. We believe that these complex phenomena may have misled the measurement of period spacing in previous studies. }

%This agreement further reinforces the reliability of our measurements.% and supports the overall consistency between the two studies.

In the bottom left panel of Fig.~\ref{fig:result_comparison}, we compare the measurements of the coupling factor ($q$) between our results and those from \cite{mosser18}. For this comparison, we selected 99 stars that are common to both studies. The comparison clearly demonstrates a correlation between our results and the previous study, with our measurements exhibiting much smaller uncertainties.

In the bottom right panel of Figure~\ref{fig:result_comparison}, we present the comparison of the g-mode phase \seb{offset} ($\varepsilon_\mathrm{g}$) between our results and those reported in \cite{mosser18}. For this comparison, we once again use the sample of 99 stars that are common to both studies. 
Unlike the previous parameters we compared, we do not observe a clear correlation between our results and the previous study in terms of the g-mode phase. This lack of correlation can be attributed to the fact that the g-mode phases are confined to a very narrow range \citep[$0.28\pm0.08$ as reported in][]{mosser18, Takata2016} and the typical uncertainties associated with the measurements are also on a similar scale. \LG{Our results of $\varepsilon_\mathrm{g}$ indeed show the same distribution. }Both our results and those from \cite{mosser18} exhibit outliers, indicating the presence of some stars that deviate from the expected g-mode phase behaviour. These outliers may potentially \seb{be related to the existence of an internal magnetic field, as mentioned in \cite{Li2022Nature}. We are currently investigating this possibility, which will be reported in a follow-up publication.} 
%point towards the existence of new and previously unidentified mechanisms operating within the cores of red giants, such as the frequency shift by central magnetic field. %Further investigations and studies are required to gain a deeper understanding of these mechanisms and their implications for stellar evolution.

\begin{figure}
    \centering
    \includegraphics[width=\linewidth]{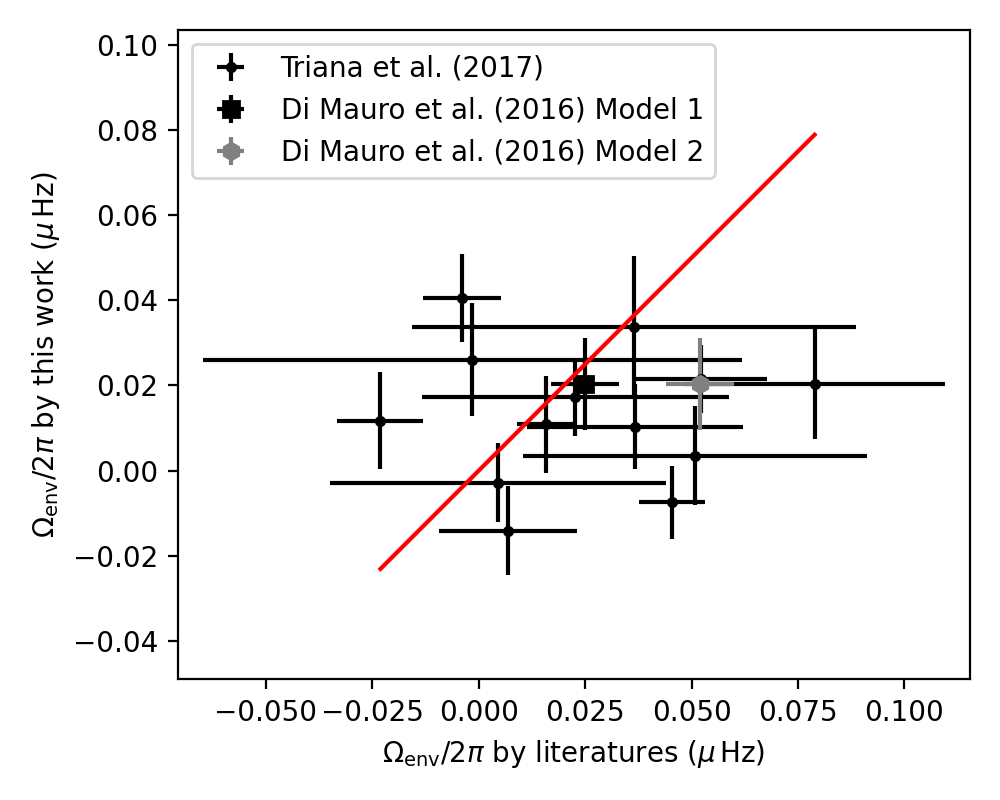}
    \caption{Comparison of envelope rotation rates between this work and literature. We collected the 13 red giant stars by \cite{Triana2017} and KIC\,4448777 by \cite{DiMauro2016}. The red line shows the 1:1 relation. }
    \label{fig:envelope_rotation_comparison}
\end{figure}

\LG{There are only a few measurements of envelope rotations available. We gathered the envelope rotation rates for 13 red giant stars from \cite{Triana2017} and for KIC\,4448777 from \cite{DiMauro2016}. \seb{We first mention that in these studies, envelope rotation rates were obtained using several different techniques (rotation inversions with several reference models, linear fits to the relation between the rotation splitting and $\zeta$), which provided results showing poor statistical agreement with one another. Rotation inversions tend to provide larger envelope rotation rates, which might be due to a contamination from the core, because the weights in the core region cannot be completely eliminated \citep[e.g.][]{DiMauro2016}. Envelope rotation measurements using $\zeta$ are generally smaller, and in some cases negative. This could point to problems related to NDE, as we showed in Sect.~\ref{subsec:Approach_comparison}. \cite{DiMauro2016} reported asymmetries in $l=1$ multiplets, which could be produced by NDE, and \cite{Deheuvels2017} have indeed shown that non-negligible NDE is expected for this star. Additionally, the targets studied by \cite{Triana2017} have p-mode large separations between about 9 and 13 $\mu$Hz, a range in which our results have shown that envelope rotation measurements can be incorrectly estimated if NDE is not taken into account. To compare our results with those of \cite{Triana2017},} 
%For the results by \cite{Triana2017}, 
we picked their envelope rotation rates determined by the linear fit between the rotational splitting and the asymptotic $\zeta$ values \citep[sixth column in Table 2 of ][]{Triana2017}, which is the closest to our approach, but does not account for NDE. For \cite{DiMauro2016}, we also picked the two values obtained from  the splitting-versus-$\zeta$ method instead of the inversion approaches. The labels `Model 1' and `Model 2' in Fig.~\ref{fig:envelope_rotation_comparison} correspond to the two best-fitting stellar evolution models that were used by \cite{DiMauro2016}. \seb{The comparison between these results and our measurements are shown in Fig.~\ref{fig:envelope_rotation_comparison}. The rather large error bars of the measurements from the literature make it difficult to see a strong correlation with our measurements. The strongest disagreements arise for stars for which \cite{Triana2017} measured negative envelope rates, which as argued above might stem from the absence of treatment of NDE in this work.}
%Although we did not find a strong correlation, it cannot determine whether our method or the literature is unreliable, as the error bars for the envelope rotation rates are of a similar magnitude to the spread of the data points. 
%We also notice that the multiple approaches by \cite{Triana2017} and \cite{DiMauro2016} yielded quite different envelope rotation rates, implying that the current methodologies for determining envelope rotation rates are still considerably uncertain, both theoretically and observationally. 
}

%\LG{In short, our results are in close agreement with the previous studies, confirming the consistency and reliability of Approach Two.}

\section{Discussion}\label{sec:discussions}

\subsection{Core rotation rate $\Omega_\mathrm{core}$}

\begin{figure*}
    \centering
    \includegraphics[width=\linewidth]{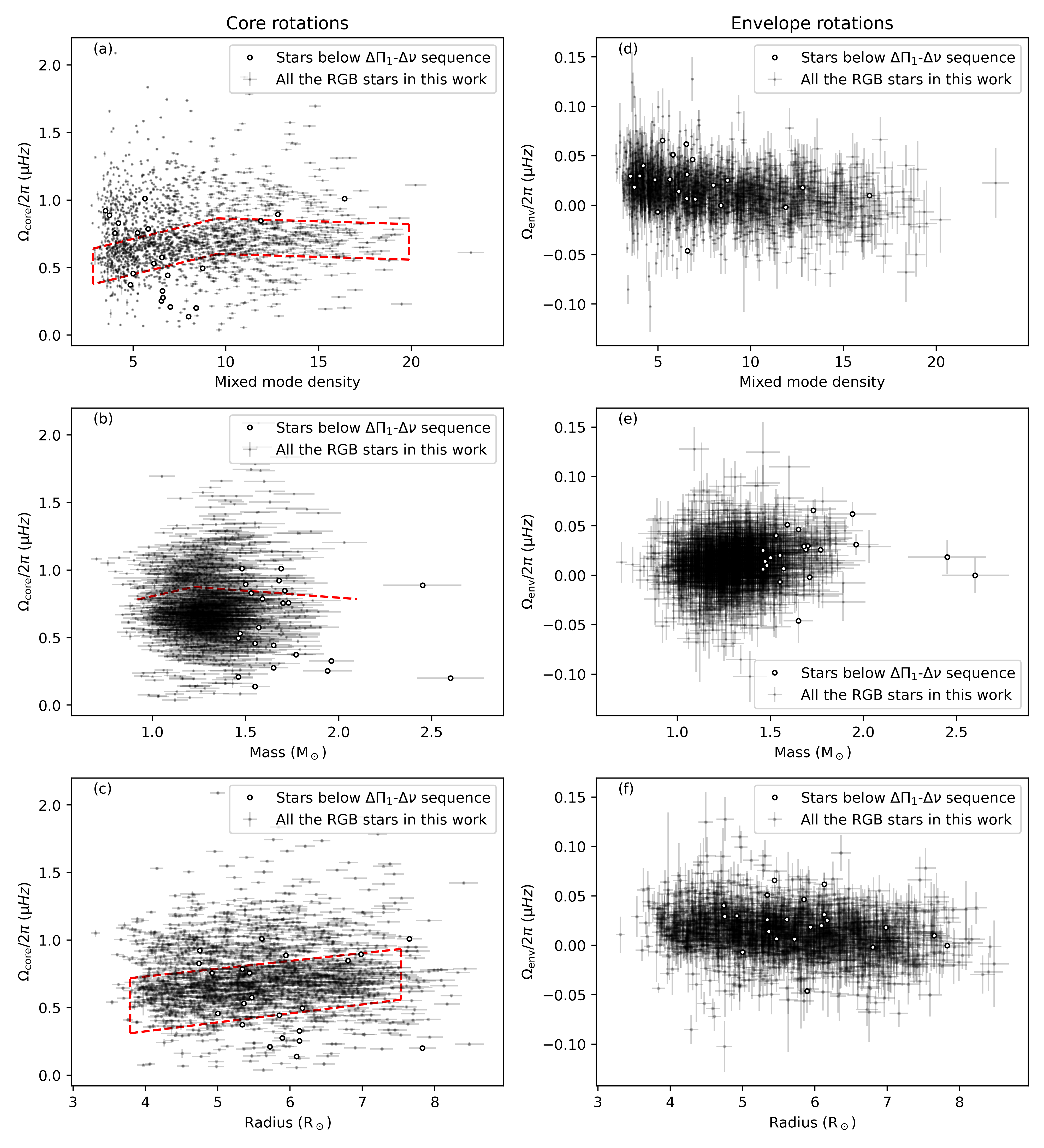}
    \caption{Results of the core and the envelope rotation rates of our sample. The left column (panel a b c) shows the core rotation rates as a function of mixed mode density, stellar mass, and radius. The right column (panel d e f) is the counterpart of envelope rotations. The open points are the stars identified below the $\Delta \Pi_1$--$\Delta \nu$ sequence in Fig.~\ref{fig:delta_nu_vs_delta_pi}. The dashed lines in panel (a) circle the over-density ridge. }
    \label{fig:rotation_both_p_and_g}
\end{figure*}

Figure~\ref{fig:rotation_both_p_and_g} displays the measurements of the core and the envelope rotation rates of \rotationnumber~stars in our sample. We show both core and envelope rotations as a function of mixed mode density, stellar mass, and radius.%, and highlight the stars below the $\Delta \Pi_1$--$\Delta \nu$ sequence as the red dots.

In panel (a), we show the core rotation rates as a function of mixed mode density, which is defined as $\mathcal{N}=\Delta \nu/\left(\Delta \Pi_1 \nu_\mathrm{max}^2\right)$ and has been shown as a good proxy of stellar evolution \citep{Gehan2018}. 
We find that the points show a \jb{bimodal} %bi-model
distribution, formed by a narrow over-density ridge and an extended background. The over-density ridge is located at approximately $\Omega_\mathrm{g}/2\pi \sim 0.6\,\mu\mathrm{Hz}$ with width $\sim\pm0.1\,\mu\mathrm{Hz}$ (circled by the dashed \LGvtwo{red} lines in panel (a) of Fig.~\ref{fig:rotation_both_p_and_g}). We visually find that the over-density ridge might slightly increase with evolution while the extended background remains consistent with evolution. Overall speaking, no clear correlation between core rotation and evolution is apparent. We give a further discussion of the over-density ridge in Section~\ref{subsec:overdensityridge}.

In panel (b) of Fig.~\ref{fig:rotation_both_p_and_g}, we present the core rotation rates as a function of stellar mass which is determined by the scaling relation proposed by \cite{Yu2018}. Many studies have shown that main sequence stars rotate nearly rigidly and their rotation rates increase rapidly with stellar mass, especially from $\sim1.0\,\mathrm{M_\odot}$ to $\sim1.4\,\mathrm{M_\odot}$, which is the transition range from stars with convective envelopes to stars with radiative envelope \citep{Royer2007, McQuillan2014, LiGang2020}. However, in red giant stars, we find no obvious correlation between core rotation rate and stellar mass. This suggests that the redistribution of angular momentum transport after the main sequence is sufficiently efficient to erase the previous rotation information. \LGvtwo{\cite{Eggenberger2022} showed that the calibrated version of the Tayler instability predicts no correlation of the core rotation rates with stellar mass for RGB stars, consistent with our observations. }
The over-density ridge mentioned in panel (a) is more significant in panel (b). We see a clear gap at $\Omega_\mathrm{core}\approx 0.88\,\mathrm{\mu Hz}$ \LGvtwo{(marked by the dashed red line).} \LGvtwo{This gap is also discovered by Hatt et al. (in prep).} 

Panel (c) presents the core rotation rate as a function of stellar radius. %jb %In panel (c), we
\jb{We} still observe the over-density ridge for core rotation rates \LGvtwo{(circled by the dashed red lines)}. The over-density ridge exhibits a slightly increasing trend between the core rotation rate and radius, indicating that as the stellar radius expands, the core experiences a slight acceleration in its rotation since the core is contracting. 

In Fig.~\ref{fig:rotation_both_p_and_g}, the open circles refer to the stars whose $\Delta \Pi_1$ are smaller than expected \citep{Deheuvels2022A&A, Rui2021MNRAS}. These stars might undergo a mass transfer or merger process, but do not show any special distribution of core rotation rates. We will provide more information in Section~\ref{subsec:below_Pi_nu_sequence}.

\subsection{The over-density ridge of $\Omega_\mathrm{core}$}\label{subsec:overdensityridge}

To further characterise the over-density ridge of the core rotations in panel (a) of Fig.~\ref{fig:rotation_both_p_and_g}, we divided the mixed mode density $\mathcal{N}$ into several bins (with $\mathcal{N} \in$ [3, 5], [5, 7], [7, 9], [9, 11], [11, 13] ), and fitted a two-Gaussian probability function to the core rotations in each bin. The two-Gaussian probability function is constructed as 
\begin{equation}
    P\left(\frac{\Omega_\mathrm{core}}{2\pi}\right) = \frac{1}{A} \exp{\left( -\frac{1}{2}\left(\frac{\frac{\Omega_\mathrm{core}}{2\pi}-\mu_1}{\sigma_1}\right)^2 \right)} + \frac{K}{A} \exp{\left( -\frac{1}{2}\left(\frac{\frac{\Omega_\mathrm{core}}{2\pi}-\mu_2}{\sigma_2}\right)^2 \right)}, \label{equ:two-gaussian_function}
\end{equation}
formed by the sum of two Gaussian distributions. The first term on the right-hand side represents the over-density ridge with a mean of $\mu_1$ and a standard deviation of $\sigma_1$, while the second term shows the extended background with a mean of $\mu_2$ and a standard deviation of $\sigma_2$. The parameter $K$ represents the ratio of the height of the two Gaussian distributions. Since the over-density ridge exhibits higher distribution density, we require $K$ to be smaller than 1 but larger than 0. Additionally, the over-density ridge also shows a smaller scatter than the background, so we set $0 < \sigma_1 < \sigma_2$. The normalisation parameter $A$ is used to ensure that the integration of the probability function equals one. 

To search for the best-fitting of $P\left(\frac{\Omega_\mathrm{core}}{2\pi}\right)$, we maximise the likelihood function defined as
\begin{equation}
    L\left( \left\{ \frac{\Omega_{\mathrm{core},i}}{2\pi} \right\} \large \Bigg| \left\{ \mu_1, \sigma_1, \mu_2, \sigma_2, K \right\} \right) \propto \prod \limits_{i} P\left(\frac{\Omega_{\mathrm{core},i}}{2\pi}\right). \label{two-gaussian-likelihood}
\end{equation}
The likelihood function $L$ represents the probability of obtaining the observed core rotation distribution $\left\{ \frac{\Omega_{\mathrm{core},i}}{2\pi} \right\}$ given the parameter set $\left\{ \mu_1, \sigma_1, \mu_2, \sigma_2, K \right\}$. It is calculated as the product of probabilities at each core rotation rate $\frac{\Omega_{\mathrm{core},i}}{2\pi}$, where $i$ means the $i^\mathrm{th}$ core rotation value. To maximise this likelihood, we also used a Markov Chain Monte Carlo (MCMC) code implemented in the Python package emcee \citep{Foreman-Mackey2013PASP}. Non-informative priors were used for the parameters.

\begin{figure*}
    \centering
    \includegraphics[width=\linewidth]{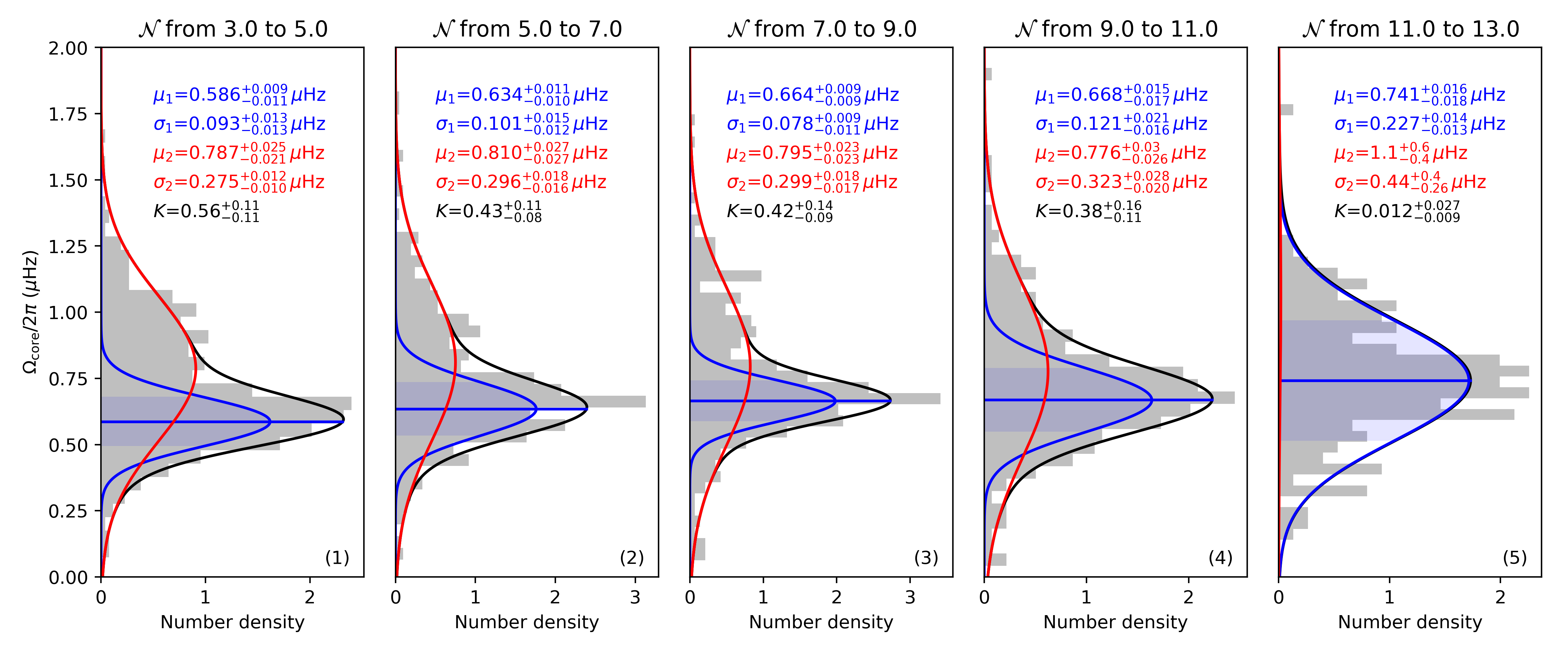}
    \caption{Two-Gaussian fit of the core rotation rate distributions in each bin of mixed mode density $\mathcal{N}$. The grey histograms are the observed distributions of core rotation rates. The blue lines correspond to the first term on the right-hand side of Eq.~\ref{equ:two-gaussian_function}, characterised by $\mu_1$ and $\sigma_1$, which represent the over-density ridge. The blue horizontal lines and shaded areas illustrate the means and 1-$\sigma$ regions of the over-density ridge in each bin. The red lines depict the second term on the right-hand side of Eq.~\ref{equ:two-gaussian_function}, characterised by $\mu_2$ and $\sigma_2$, representing the extended background. The black lines indicate the sum of the two Gaussian distributions, which provide a good reproduction of the observed histogram. }
    \label{fig:two_gaussian_fit}
\end{figure*}

Figure~\ref{fig:two_gaussian_fit} presents the best-fitting two-Gaussian distributions in each bin of mixed mode density $\mathcal{N}$. We find that the two-Gaussian distributions effectively reproduce the observed histograms of core rotation rates when $\mathcal{N}$ ranges from 3 to 11 (panels (1) to (4)), as indicated by the well-constrained parameters obtained from the MCMC fit. However, for $\mathcal{N}$ values from 11 to 13 (panel (5)), we face challenges in obtaining accurate constraints for $\mu_2$, $\sigma_2$, and $K$. Particularly, the MCMC chains suggest that $K$ converges to zero, suggesting that a single Gaussian distribution adequately reproduces the observed histogram. 

\LGvtwo{
To further confirm whether the two-Gaussian model (Eq.~\ref{equ:two-gaussian_function}, hereafter named model $M_2$) is statistically more significant than a single-Gaussian model (model $M_1$) to represent the distribution of core rotation rates, we computed the odd ratio, that can be reduced to the Bayes factor \citep[e.g.][]{Deheuvels2015} defined as
\begin{equation}
    B_{21}=\frac{P(D|M_2)}{P(D|M_1)},
\end{equation}
where $P(D|M_i)$ is the global likelihood (or evidence) of model $M_i$, that is the probability for model $M_i$ to produce the data set $D$. It is obtained by marginalisation, that is the integration of the likelihood over the whole parameter space \citep[see discussion for example in][]{Benomar2009}. To compute the evidences of models $M_1$ and $M_2$, we use a parallel tempering MCMC code we developed in the past \citep{Deheuvels2015}. We then computed the logarithmic Bayes factor $\ln B_{21}$ to test the ability for model $M_2$ to better represent the distribution of rotation than model $M_1$. For the four distributions obtained with ${\cal N} < 11$, we get $\ln B_{21}$ ranging between 29 and 13. According to Jeffreys' scale \citep{Jeffreys1961}, values above 5 are decisive evidences for model $M_2$ against model $M_1$. However, for the last distribution ($11<{\cal N}<13$), we get $\ln B_{21} \approx -0.5$, that confirms that a single Gaussian law is enough to model this distribution.
Therefore, we confirm that the over-density ridge is statistically significant up to $\mathcal{N} = 11$. 
%To further confirm whether the two-Gaussian model is statistically more significant than the single-Gaussian model, we calculated the odds ratio \citep[e.g.][]{Deheuvels2015}, which is defined as the probability ratio of the two models,
%\begin{equation}
%    O_{12} = L_1 / L_2.
%\end{equation}
%In the equation, subscript `1' denotes the single-Gaussian model, and subscript `2' denotes the two-Gaussian distribution model. $L_1$ and $L_2$ are the probability of reproducing the observed data using the model and given parameters defined by Eq.~\ref{two-gaussian-likelihood}. To calculate $L_1$, we just forced $K$ to be zero (in Eq.~\ref{equ:two-gaussian_function}). We find that the odds ratio is close to zero for each bin when $\mathcal{N}<11$, indicating that the two-Gaussian model follows the observed distribution much better than the single-Gaussian model. However, for the bin where $11 < \mathcal{N}<13$, the odds ratio is not close to zero, suggesting that the single-Gaussian model is sufficient to reproduce the observed distribution. Therefore, we confirm that the over-density ridge statistically exists only up to $\mathcal{N} = 11$ and gradually diminishes beyond this mixed-mode density. 
}

The mean value $\mu_2$ and the standard deviation $\sigma_2$ of the extended background exhibit consistency across the range of $\mathcal{N}$ values from 3 to 11, indicating no significant variations in the distributions of the extended background. On the other hand, the mean value $\mu_1$ of the over-density ridge increases from $\mathcal{N}=3$ to 7 and remains constant thereafter. This measurement of $\mu_1$ aligns with the visual inspection of the over-density ridge depicted in panel (a) of Fig.~\ref{fig:rotation_both_p_and_g}. %To illustrate this agreement, we plot the 1-$\sigma_1$ regions of the first term in Eq.~\ref{equ:two-gaussian_function} in Fig.~\ref{fig:core-rotation-with-cluster-members}. It is worth noting that the 1-$\sigma_1$ regions closely encircle the over-density ridge identified visually. Furthermore, the standard deviation $\sigma_1$ reaches its minimum value between $\mathcal{N}=7$ and 9, while it remains nearly unchanged in other bins with $\mathcal{N} < 11$.

Currently, \jb{the origin of this bimodal distribution still have to be understood.}
We investigated the stellar parameters of the stars within the over-density ridge, such as the seismic parameters ($q$, $f_\mathrm{shift}$, and $\Omega_\mathrm{core}$) or global parameters (mass, metallicity, and $\mathrm{log}g$), \jb{without finding any significant correlations or trends.}. As the bimodal distribution emerges when $\mathcal{N}< 13$, we speculate that the mechanism responsible for the formation of the bimodal distribution might be significant in late subgiant to young RGB stars, but gradually disppears with stellar evolution.

%\begin{figure}
%    \centering
%    \includegraphics[width=\linewidth]%{figures/cluster_member_core_rotation.png}
%    \caption{Core rotation rates with mixed mode density. The error bars show the 1-$\sigma$ regions of the first term on the right-hand side of Eq.~\ref{equ:two-gaussian_function}, which exhibits the location of the over-density ridge. The coloured symbols mark the RGB stars from the three clusters. }
%    \label{fig:core-rotation-with-cluster-members}
%\end{figure}

\subsection{Envelope rotations}\label{subsec:envelope_rotations}

\begin{figure}
    \centering
    \includegraphics[width=\linewidth]{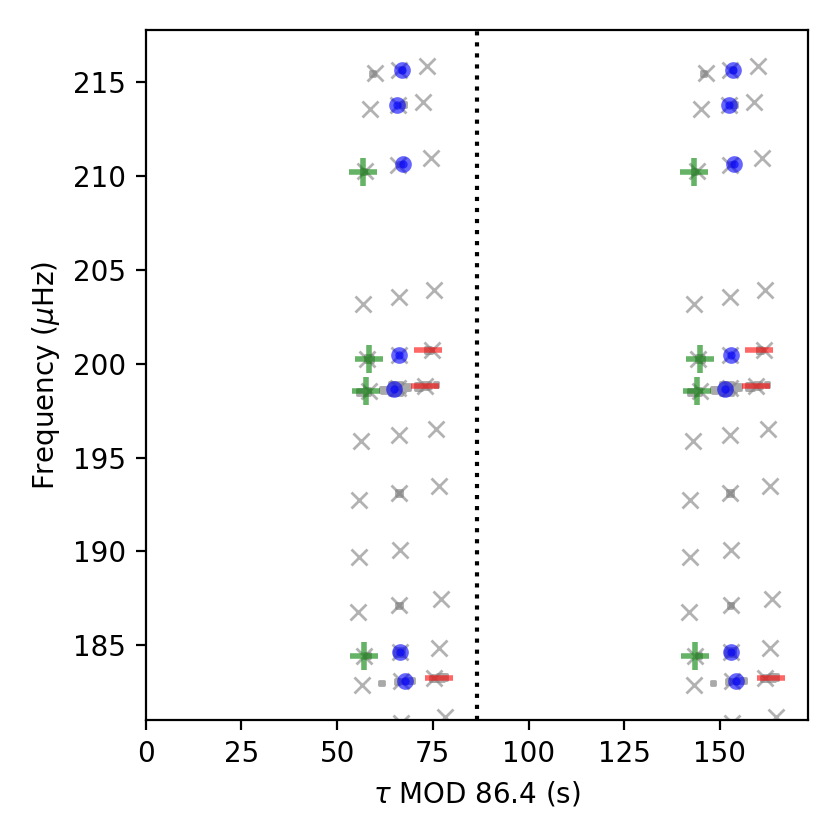}
    \caption{Stretched \'{e}chelle diagram of KIC\,10817031, which shows a negative envelope rotation rate due to the limited number of rotational splittings. }
    \label{fig:stretched_echelle_diagram_KIC10817031}
\end{figure}

Panels (d) and (e) \jb{of Fig.~\ref{fig:rotation_both_p_and_g}} show the envelope rotations as a function of mixed mode density and stellar mass. Although our code reports some negative envelope rotation rates of some stars, the envelope rotations of the majority are \jb{consistent with} %equivalent to 
zero within \jb{3 $\sigma$}. %the 3-sigma region. 
%\threesigmapositiverotationnumber~stars show envelope rotation rates that are larger than zero beyond three times the standard deviation, and only 33 stars show negative envelope rotation rates outside the 3-sigma region. 
\jb{We detect significant (above a 3-$\sigma$ detection threshold) positive envelope rotation rates  in \threesigmapositiverotationnumber~stars. We also measure negative rotation rates above a 3-$\sigma$ detection threshold in only 33 stars.}
The reason for the negative values is twofold. First, most of the stars in our sample have envelope rotation rates that are distributed around 0.025\,$\mathrm{\mu Hz}$ (corresponding to 463 days), which is close to the detection limit of 1400-day asteroseismic data by \emph{Kepler}. Second, p-dominated mixed modes usually have wider linewidths and larger frequency uncertainties, which results in fewer constraints on envelope rotation compared to g-dominated modes on core rotation. Consequently, a few negative envelope rotation results might be reported by our code. \LG{In Fig.~\ref{fig:stretched_echelle_diagram_KIC10817031}, we show the stretched \'{e}chelle diagram of KIC\,10817031, which exhibits the smallest negative value of envelope rotation ($\Omega_\mathrm{env}/(2\pi)=-0.102\pm0.026\,\mathrm{\mu Hz}$) among our sample. We attribute the negative envelope rotation in this star to the limited number of splittings observed in p-dominated modes, which do not provide sufficient constraints on the envelope rotation rate. }

In panel (d), a significant decreasing trend is observed between envelope rotation and mixed mode density. 
This can be attributed to the expansion of the stellar envelope, which leads to a deceleration of the envelope rotation (this interpretation is further supported by panel (f), which confirms that the envelope rotation rate shows a decreasing trend as a function of the stellar radius). 
%primarily caused by the expansion of stellar radii during the red giant phase. 
At the beginning of the red giant phase (with a mixed mode density around 4), the envelope rotation rates are distributed between 0 and 0.05\,$\mu\mathrm{Hz}$, corresponding to rotation periods of approximately 460 days. As stars expand, the envelope rotation rates decrease and eventually distribute around 0\,$\mu \mathrm{Hz}$. This indicates that current asteroseismic data provide limited constraints on envelope rotation rates due to the expansion of stars. In panel (e), the relation between envelope rotation rates and stellar masses is examined, yet no correlation is found between them.

%Panel (f) displays the envelope rotation rate as a function of radius. The envelope rotation rate shows a decreasing trend with stellar radius. This can be attributed to the expansion of the stellar envelope, which leads to a deceleration of the envelope rotation. 

%In Fig.\ref{fig:envelope_rotation_vs_radius}, we examine the relation between envelope rotation and stellar radius. We consider the red-giant-branch (RGB) stars from our study, specifically those with envelope rotation rates that are larger than zero beyond 3-sigma region. 
% This selection criteria results in a total of \threesigmapositiverotationnumber~RGB stars plotted in Fig.\ref{fig:envelope_rotation_vs_radius}. 

\begin{figure}
    \centering
    \includegraphics[width=\linewidth]{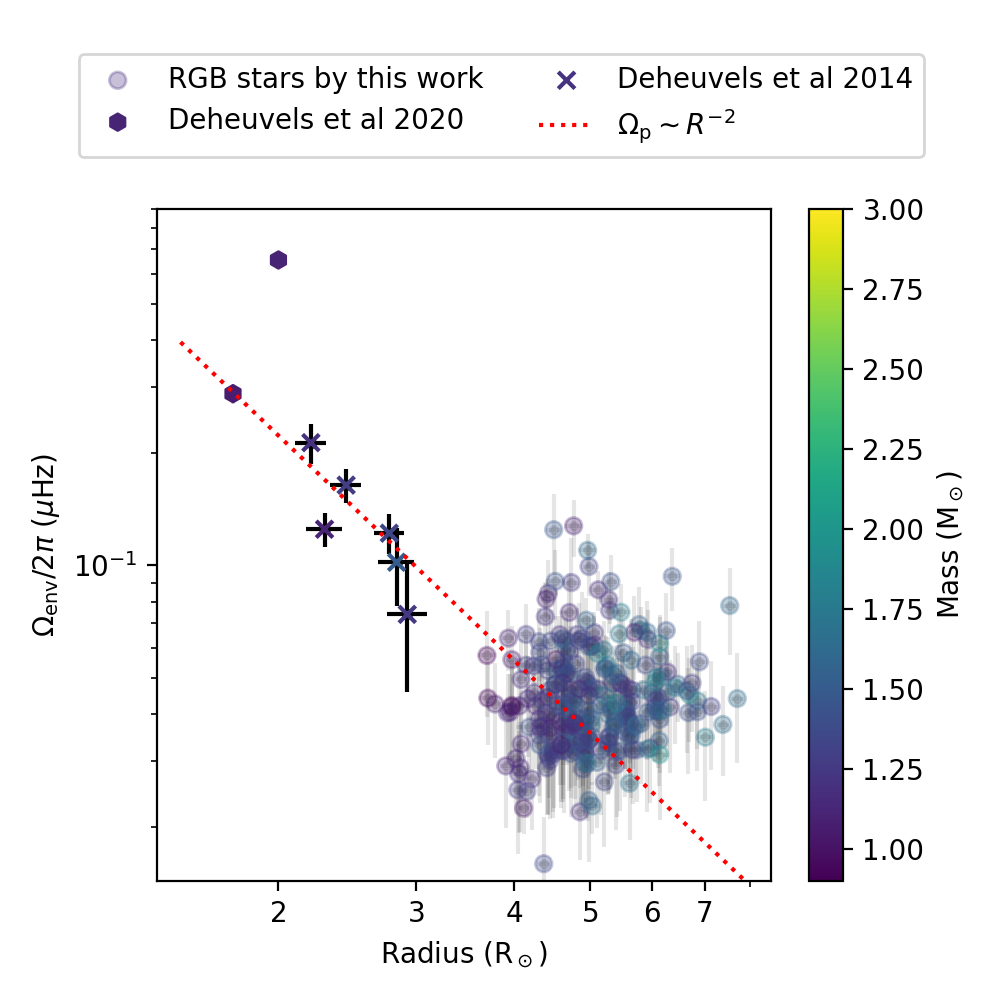}
    \caption{Envelope rotation rate as a function of stellar radius. We include \threesigmapositiverotationnumber~RGB stars by this work whose envelope rotation rates are outside three sigma deviation from zero, six young RGB stars by \cite{Deheuvels2014}, and two young RGB stars by \cite{Deheuvels2020}. The red dotted line shows the best fitting results assuming envelope rotation decreases with $R^{-2}$, which is $\Omega_\mathrm{p}/2\pi = 0.97/R^2$.}
    \label{fig:envelope_rotation_vs_radius}
\end{figure}

In Fig.\ref{fig:envelope_rotation_vs_radius}, \jb{we plot the relation between envelope rotation $\Omega_{\rm env}$ and stellar radius $R$ for the \threesigmapositiverotationnumber~RGB stars with significant measurements.}
Additionally, we include data from other studies, including six young post-main-sequence stars from \cite{Deheuvels2014}, and two young subgiants from \cite{Deheuvels2020}. In a simplified scenario, the envelope rotation is expected to %follow a relation with $R^{-2}$, where $R$ represents the stellar radius. 
\jb{scale as $R^{-2}$.}
We find that this relation holds for the RGB stars in our sample. The best-fitting relation is given by
\begin{equation}
    \Omega_\mathrm{env}/2\pi~(\mu\mathrm{Hz}) = 0.89/R^2~(\mathrm{R_\odot}),
\end{equation}
plotted as the red dotted line in Fig.~\ref{fig:envelope_rotation_vs_radius}. Notably, the young RGB stars with $R<3.5\,\mathrm{R_\odot}$ exhibit a smaller scatter around this relation. In contrast, the RGB stars from our study with $4\,\mathrm{R_\odot} < R < 8 \,\mathrm{R_\odot}$ show a larger spread in their envelope rotation rates. \LGvtwo{There may be several reasons for the increased spread of envelope rotation rates during the transition from subgiants to young red giants. One reason is the bias of the data; currently, there are only eight subgiant samples \citep{Deheuvels2014, Deheuvels2020}, which is far from sufficient to delineate the statistical characteristics of surface rotations in subgiants. Another reason is that as the stars expand, certain physical processes, such as the engulfment of planets \citep{Tayar2022, Kunitomo2011ApJ, Hon2023Natur} or the tidal effects of nearby companions \citep[e.g.]{Ahuir2021}, become more significant. These can all potentially impact the envelope rotation rates.}

The stars which might undergo a mass transfer or merger process (open circles in Fig.~\ref{fig:rotation_both_p_and_g}) still do not show any special envelope rotation rates, implying that the mass transfer or merger process might not lead to an obvious change in surface rotation rates.
%Furthermore, we observe that the envelope rotations of helium-burning stars by \citet{Deheuvels2015} (not shown in Fig.\ref{fig:envelope_rotation_vs_radius}) 
%Furthermore, we \jb{also compared this trend with the envelope rotations measured in seven helium-burning stars by \citet{Deheuvels2015} (not shown in Fig.\ref{fig:envelope_rotation_vs_radius}) and noticed that they} deviate from the predicted $R^{-2}$ decay, as they rotate much faster. This discrepancy may suggest the presence of an efficient mechanism for angular momentum transport during the transition from RGB to \jb{red clump}. %RC.
%Alternatively, it could be attributed to the fact that the helium-burning stars reported by \cite{Deheuvels2015} have significantly larger stellar masses compared to our sample.

\subsection{Differential rotation}\label{subsec:differential_rotation}

\begin{figure*}
    \centering
    \includegraphics[width=\linewidth]{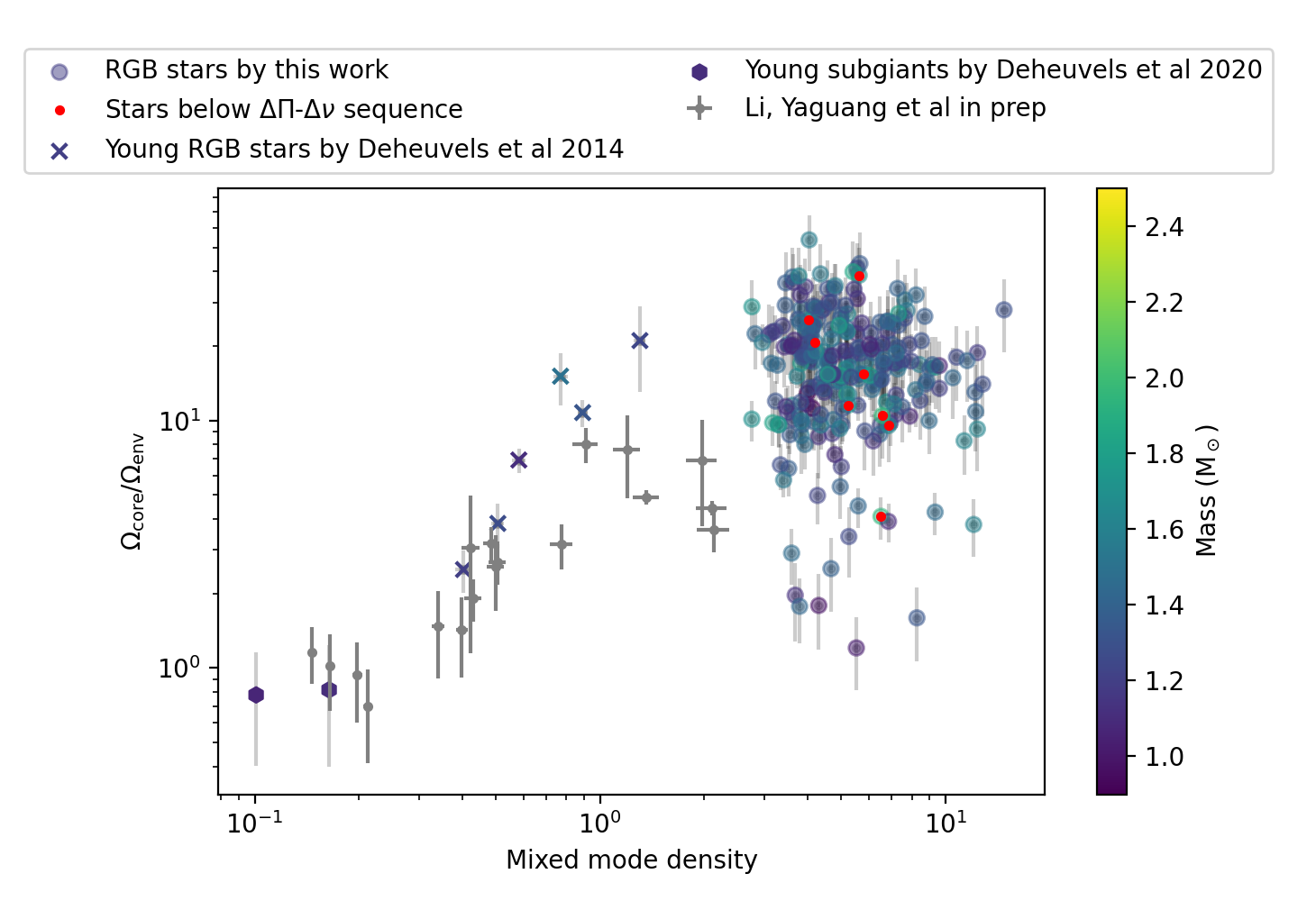}
    \caption{The ratio between core and envelope rotation rates of evolved stars, including \threesigmapositiverotationnumber~RGB stars by this work whose envelope rotation rates are outside three sigma deviation from zero. We also show the six young RGB stars by \cite{Deheuvels2014} and seven helium-burning stars by \cite{Deheuvels2015}. }
    \label{fig:differential_rotation_vs_mixed_mode_density}
\end{figure*}

Our analysis of core and envelope rotation rates in RGB stars allows us to examine the core-to-surface differential rotations, specifically the ratio between the core and envelope rotation rates, denoted as $\Omega_\mathrm{core}/\Omega_\mathrm{env}$. %To ensure reliable measurements and avoid issues associated with near-zero or negative estimates of $\Omega_\mathrm{env}$, we adopt a selection criterion where only stars with envelope rotation rates that are three times the standard deviation larger than zero are considered (the same as the \threesigmapositiverotationnumber~stars in Fig.~\ref{fig:envelope_rotation_vs_radius}). 
To ensure reliable measurements and avoid issues associated with near-zero or negative estimates of $\Omega_\mathrm{env}$, we \jb{ only consider the \threesigmapositiverotationnumber~stars with a significant measurements in the envelop (the same stars as in Fig.~\ref{fig:envelope_rotation_vs_radius}).}
It is important to note that this selection %criterion 
leads to a \LG{bias} 
%of the reported ratio, 
\jb{towards low values of the reported ratio distribution,}
as many stars with near-zero envelope rotations are discarded. 

In Figure~\ref{fig:differential_rotation_vs_mixed_mode_density}, we examine the relation between the ratio of core-to-envelope rotation rates, $\Omega_\mathrm{core}/\Omega_\mathrm{env}$, and the mixed mode density $\mathcal{N}$. This analysis provides insights into how differential rotation evolves with stellar evolution. In addition to the \threesigmapositiverotationnumber~red-giant-branch (RGB) stars analysed in this work, we also include six young red giants studied by \cite{Deheuvels2014}, two young subgiants exhibiting solid-body rotation profiles reported by \cite{Deheuvels2020}, and 17 young red giants by Li, Yaguang et al. (in prep). 
We find that as stars evolve, the ratio $\Omega_\mathrm{core}/\Omega_\mathrm{env}$ increases significantly from around 1 at $\mathcal{N}\sim 0.1$ to approximately 20 at $\mathcal{N}\sim 1$ \citep{Deheuvels2014, Deheuvels2020}. Our sample consists of more evolved stars, with $\mathcal{N}$ ranging from approximately 3 to 20. The core-to-envelope rotation ratios of our stars fall within the range of around 10 to 50, exhibiting a larger spread compared to the young stars. \LGvtwo{We speculate that this increased spread may be caused by the increased spread of envelope rotation rates (we refer to the discussion in Sect.~\ref{subsec:envelope_rotations}).} Moreover, based on the measurements of the seven helium-burning stars reported by \cite{Deheuvels2015}, Once the stars enter the helium-burning phase, their differential rotations become mild, and the core-to-envelope rotation ratios are typically around two or three.

It is intriguing to note that six stars in our sample exhibit extremely mild differential rotations, with $\Omega_\mathrm{core}/\Omega_\mathrm{env}<2$. They are KIC\,4279009, 6956834, 7257241, 7630743, 8352953, and 11294612. Among them, KIC\,7630743 is the one that has the smallest $\Omega_\mathrm{core}/\Omega_\mathrm{env}$. Our measurements indicate that $\Omega_\mathrm{core}/2\pi = 0.065\pm0.007\,\mathrm{\mu Hz}$ and $\Omega_\mathrm{env}/2\pi = 0.063\pm0.012\,\mathrm{\mu Hz}$, hence the ratio $\Omega_\mathrm{core}/\Omega_\mathrm{env}$ is only $1.04\pm0.24$. KIC\,7630743 thus becomes the first red giant that is found to have a nearly uniform rotation. We show the rotational splittings of KIC\,7630743 in Fig.~\ref{fig:splittings_of_7630743}. 

\LG{We find that these rigidly rotating RGB stars exhibit typical envelope rotation rates. Their rigid rotation profiles are attributed to the extremely slow rotation rates of their cores. One hypothesis is that their progenitors, during the main sequence phase, were extremely slow rotators within binary systems, as discovered by \cite{LiGang2020binary}. These stars exhibit $\gamma$\,Doradus-type pulsations, and their near-core rotation rates are on the order of hundreds of days. \cite{Fuller2021_inverse_tides} proposed the concept of "inverse tides" to explain this slow rotation: tidal interaction with unstable pulsation modes can force the spins of the stars away from synchronicity. As these stars evolve into the RGB phase, they may retain their slowly rotating cores, resulting in rigid rotation profiles.
}

\subsection{Stars below $\Delta\Pi_1$--$\Delta\nu$ degenerate sequence and their $q$ values}\label{subsec:below_Pi_nu_sequence}
\begin{figure}
    \centering
    \includegraphics[width=\linewidth]{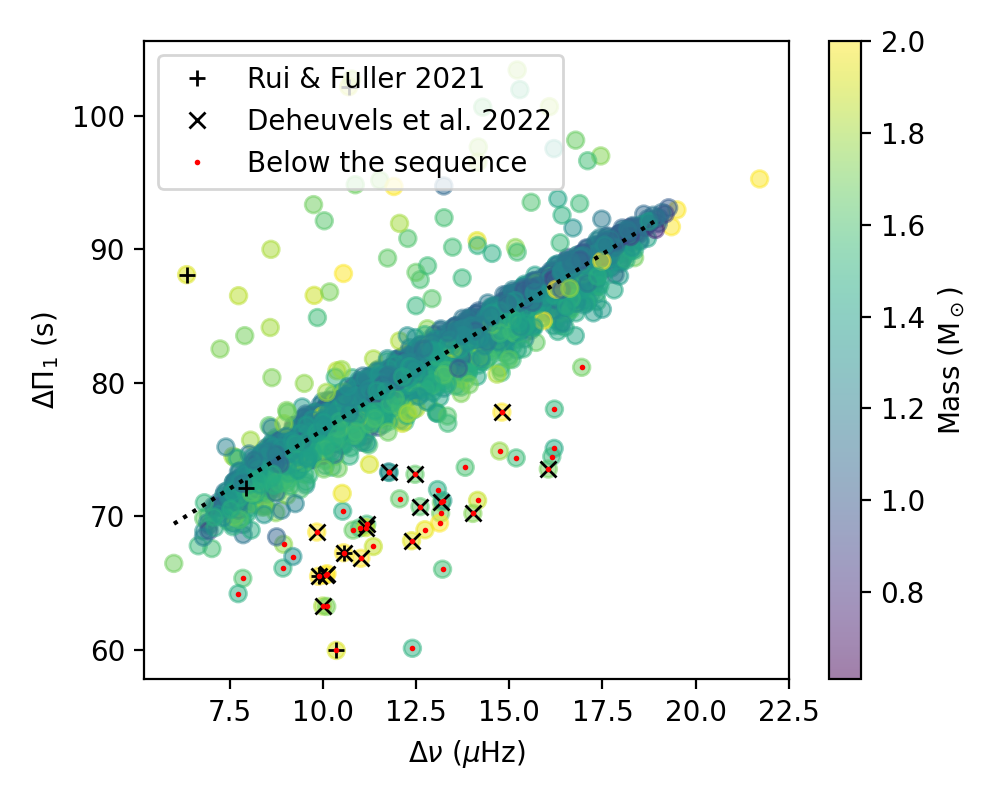}
    \caption{$\Delta\Pi_1$--$\Delta\nu$ degenerate sequence, colour-coded by the scaling-relation mass by \cite{Yu2018}. The plus symbol (`+') marks the stars reported by \cite{Rui2021MNRAS}, and the cross symbol (`$\times$') are reported by \cite{Deheuvels2022A&A}. The red dots mark the stars that are located six seconds below the sequence (dotted line) by this work. }
    \label{fig:delta_nu_vs_delta_pi}
\end{figure}

\LGnew{In this section, besides the rotation part, we report the result of the $\Delta\Pi_1$--$\Delta\nu$ sequence. }The $\Delta\Pi_1$--$\Delta\nu$ sequence was noticed by several previous studies \citep[e.g.][]{Stello2013ApJ, Mosser2014A&A, Vrard2016A&A}. \citet{Deheuvels2022A&A} demonstrated that red giants join the $\Delta\Pi_1$--$\Delta\nu$ sequence when electron degeneracy becomes strong in the cores. \LGvtwo{The stars above the sequence might be because their cores have not yet reached electron degeneracy, or their masses are too large to reach the degeneracy. For those below the sequence, a process of mass transfer or a stellar merger was proposed to address the core-to-envelope inconsistency \citep[see][]{Deheuvels2022A&A, Rui2021MNRAS}.} \LG{Strong central magnetic fields can also generate smaller apparent $\Delta \Pi_1$ \citep{Loi2020, Li2022Nature, Bugnet2022, Deheuvels2023}, but we have removed them from our sample.} Here we present our measurements of the degenerate sequence in Fig.~\ref{fig:delta_nu_vs_delta_pi}. Our results successfully reproduce the sequence, and we also observe the expected mass gradient, with higher-mass stars tending to be located at the lower boundary of the sequence. We show the mass-transfer or merger remnants by \cite{Deheuvels2022A&A} and \cite{Rui2021MNRAS}, and we also identify more stars that are located below the sequence (marked by the red dots). The criterion is that the $\Delta\Pi_1$ value of a star is six seconds smaller than the $\Delta\Pi_1$-$\Delta\nu$ linear fit. 

\jb{In order to test further the scenario that these stars have undergone a mass transfer process,}
%Since these stars located below the $\Delta \Pi_1$-$\Delta\nu$ are considered to have undergone a mass transfer process, 
we examined whether they exhibit any unusual parameter distributions. \jbnew{These stars are highlighted in Fig.~\ref{fig:rotation_both_p_and_g} as empty dots. They show normal core and envelope rotation rates.} \LGvtwo{Interestingly, as shown in panel (b) in Fig.~\ref{fig:rotation_both_p_and_g}, the stars below the $\Delta \Pi_1$--$\Delta \nu$ sequence all show relatively large masses ($\gtrsim 1.5\,\mathrm{M_\odot}$), which is another evidence that they underwent a process of mass transfer or merger, but these stars still show normal core rotation rates.} We \jbnew{also} examined their \jbnew{other} parameters ($q$, $\varepsilon_\mathrm{g}$, $f_\mathrm{shift}$) and found no significant deviations, except for $q$.

\begin{figure}
    \centering
    \includegraphics[width=\linewidth]{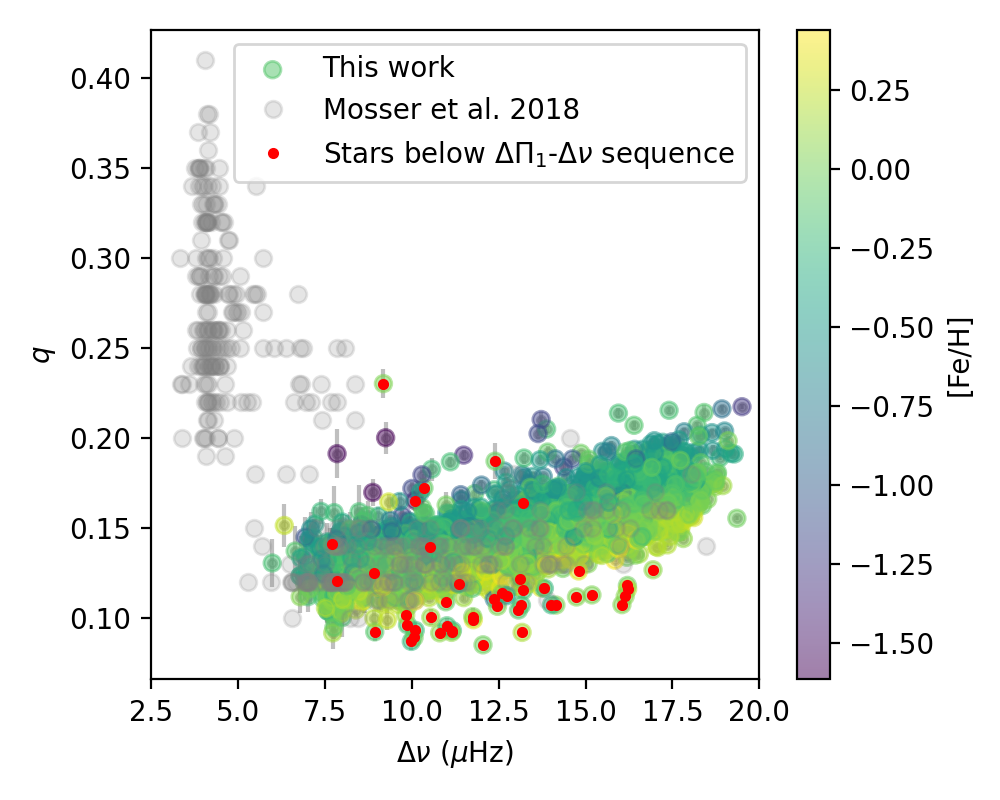}
    \caption{The coupling factor $q$ as a function of $\Delta\nu$, colour-coded by metallicity. The grey circles are reported by \cite{mosser18}. Our results are consistent with the results of the RGB stars by \cite{mosser18}, while plenty of helium-burning stars show much larger $q$ and much smaller $\Delta \nu$. A clear metallicity gradient is seen in the figure. The stars which might undergo mass transfer processes tend to have smaller $q$ value. }
    \label{fig:q_result_FeH}
\end{figure}

The coupling factor $q$ stands for the coupling strength between p- and g-mode cavities and provides structure information of the evanescent region between the two cavities for low-luminosity red giants \citep{Takata2016, Mosser2017, Hekker2018, Pincon2020,  Jiang2020}. In our study, we assume that the $q$ value is the same for all modes, although \cite{Jiang2020} has shown a frequency dependence. 
We present the measurement of the coupling factor $q$ in Fig.~\ref{fig:q_result_FeH}. We find that the value $q$ distributes between $\sim 0.2$ and $\sim 0.1$ and decreases with descending $\Delta\nu$, forming a $q$-$\Delta \nu$ linear relation \citep{Mosser2017}. The linear relation is caused by the progressive migration of the evanescent region from the radiative to the convective zones within the star \citep{Pincon2020}. In Fig.~\ref{fig:q_result_FeH}, we find that most of the stars under the $\Delta \Pi_1$-$\Delta \nu$ sequence show smaller values of $q$. This is because these stars should have older cores with smaller $q$, while mass transfer from their companions leads to an increase in mean density, resulting in a larger $\Delta \nu$. As a result, they are located below the $q$-$\Delta \nu$ sequence.

\begin{figure}
    \centering
    \includegraphics[width=\linewidth]{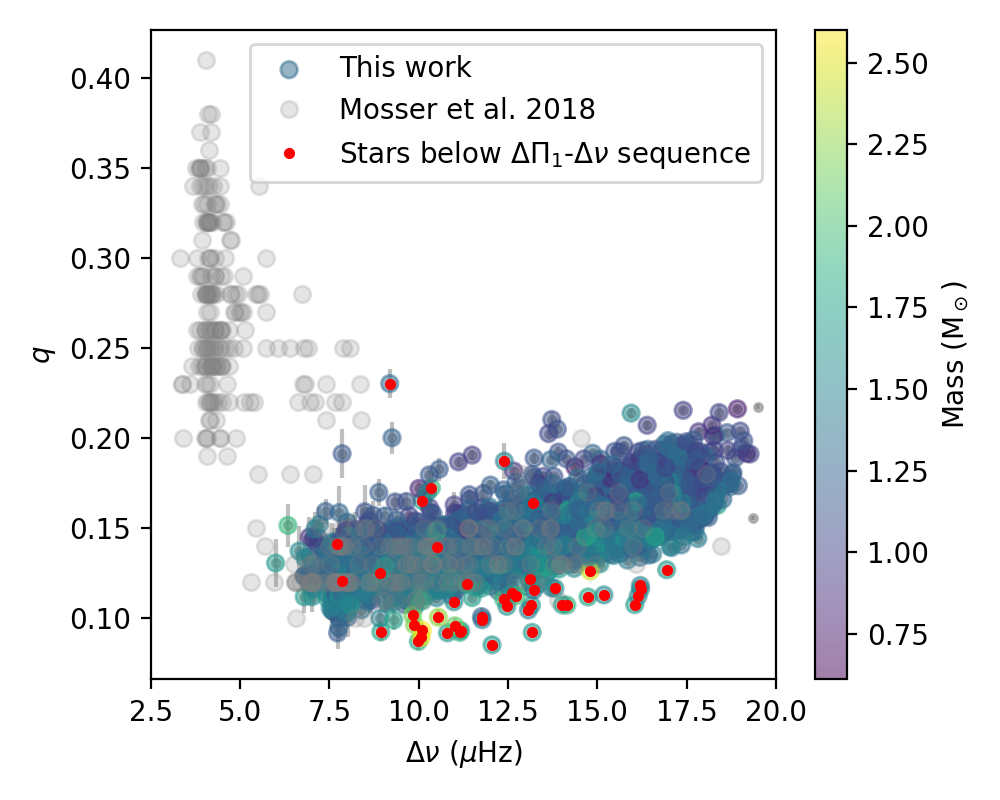}
    \caption{Same as fig.~\ref{fig:q_result_FeH} but colour-coded by stellar mass. The gradient caused by mass is weaker but is still visible. }
    \label{fig:q_result_mass}
\end{figure}

A large spread of the correlation between $q$ and $\Delta \nu$ is seen, which is caused by different metallicity and stellar mass. %As shown in Fig.~\ref{fig:q_result_FeH}, a clear metallicity gradient is evident, indicating that low-metallicity stars tend to exhibit larger $q$ values. 
\jb{The clear metallicity gradient visible in Fig.~\ref{fig:q_result_FeH} indicates that low-metallicity stars tend to exhibit larger $q$ values.}
Additionally, in Fig.~\ref{fig:q_result_mass}, we can see a weak correlation with stellar mass, although not as pronounced as the effect of metallicity. Stars with smaller masses tend to display larger values of $q$. Furthermore, we also notice that the stars below the $\Delta\Pi_1$-$\Delta \nu$ sequence show larger stellar mass ($M\gtrsim1.5\,\mathrm{M_\odot}$, also shown later in panel (b) and (e) of Fig.~\ref{fig:rotation_both_p_and_g}), consistent with the hypothesis that they gain extra material from companion stars. \LG{The dependences between $q$ and metallicity (and mass) have been reported by \cite{Kuszlewicz2023ApJ}. Our larger sample further confirms their discovery.}

\section{Conclusions}\label{sec:conclusion}

In this work, we report the largest sample of \emph{Kepler} RGB stars \LGnew{with clearly identified and analysed mixed-mode patterns. }
\sebnew{Overall, we measured mixed-mode properties for \totalstarnumber~RGB stars.}
%Finally, we collect} \totalstarnumber~such RGB stars from the \emph{Kepler} data. 
Among them, \rotationnumber~stars show rotational splittings, including \doubletnumber~stars showing doublets and \tripletnumber~stars showing triplets. There are also \norotationnumber~stars that show only $m=0$ modes. 

\sebnew{One of our main goals was to reliably measure envelope rotation rate for a large sample of RGB stars. In this work, we argued that the methods that were previously applied to infer the internal rotation of red giants (rotation inversions, linear regression of the relation between the rotational splitting and the $\zeta$ parameter) do not account for near-degeneracy effects (NDE, \citealt{Deheuvels2017}). We showed that this leads to biases in the estimates of envelope rotation rates, which become stronger as stars ascend the RGB. We proposed an alternate approach to fitting asymptotic expressions of mixed modes to the observations without resorting to the $\zeta$ function and we showed that it properly treats NDE. This new approach provides more reliable measurements of the envelope rotation rate.}
%\LGnew{We applied a new approach (Approach Two) to fit the asymptotic relation to the mixed modes. The advantage of the new approach is that it avoids using the $\zeta$ function and can take the NDE into account.} We show that the previous asymptotic expression with the $\zeta$ function may lead to many negative results of $\Omega_\mathrm{env}$ in the case of strong NDE, and our new approach indeed reports a more reliable measurement of $\Omega_\mathrm{env}$. \LGnew{Beyond the asymptotic relation, the inversion method may overestimate the envelope rotation rate, because there is inevitable pollution from the g-mode cavity that is rotating rapidly. } 

\LGnew{Using this new approach, }we measured the following mixed-mode parameters for each star: period spacing $\Delta\Pi_1$, coupling factor $q$, g-mode offset term $\varepsilon_\mathrm{g}$, frequency correction of pure dipole p mode $f_\mathrm{shift}$ \LG{(linked to the small separation $\delta \nu_{01}$)}, the mean core rotation rate $\Omega_\mathrm{core}$, and the mean envelope rotation rate $\Omega_\mathrm{env}$. \LG{Asymptotic relations were also fitted to $l=0$ and $2$ pressure mode frequencies.} The envelope rotation rate, in particular, is measured in such a large sample for the first time.

\LGnew{The core rotation sample we reported is twice as large as that in previous studies, allowing us to investigate the detailed structure of core rotation as it evolves. }We confirm previous findings that there is no obvious correlation between the core rotation rates and evolution. \LGnew{However, we uncovered the existence of an over-density ridge, or bimodal distribution, 
%is noticed 
for core rotation rates as a function of %mixed-mode density, 
the evolution along the RGB. \sebnew{Indeed, a population of red giants shows core rotation rates that are narrowly distributed around about 0.6\,$\mu$Hz, while other red giants show larger scatter. The central peak of this newly identified distribution slightly increases with evolution, which suggests that the cores of this sub-population of red giants might be spinning up, in contrast with other red giants.}
This implies that there might be two populations of RGB stars whose cores undergo different rotational evolution. Currently, we cannot provide any theoretical explanation for the formation of the over-density ridge. However, it could provide precious information about the way angular momentum is transported in the subgiant and RGB phases.}

\LGnew{We increased the sample of the envelope rotation rate measurements by two orders of magnitude. }We find that there is a clear spin-down of the envelope rotation rates as a function of stellar evolution and radius due to the expansion of stars. \sebnew{For more evolved red giants, 
%We acknowledge that 
the envelope rotation rates of most stars are too slow to be significantly measured, even with the longest \emph{Kepler} data sets.} \LGnew{We obtain \threesigmapositiverotationnumber~stars whose envelope rotation rates are significantly larger than zero, among a total of \rotationnumber~stars for which the internal rotation could be probed. The core-to-envelope rotation ratios of these stars are measured. We find that the ratios distribute around $\sim20$. %which might be underestimated 
\sebnew{This is to be understood as the lower-end of the distribution }because we ignore a lot of stars whose envelope rotation rates are close to zero. The core-to-envelope ratios $\Omega_\mathrm{core}/\Omega_\mathrm{env}$ show a larger spread compared to the stars in the subgiant phase \citep{Deheuvels2014, Deheuvels2020}, suggesting various processes of angular momentum transport at the transition between subgiants and red giants. Our observations will be important to measure the efficiency of angular momentum transport along the RGB since for this purpose a core-envelope rotation ratio is needed \citep{Eggenberger2019I}. }Interestingly, we also find several RGB stars showing extremely mild radial differential rotations (with $\Omega_\mathrm{core}/\Omega_\mathrm{env}<2$), \sebnew{which had not been reported before. We find that the envelope rotation rates of these stars are quite typical, but their cores are rotating much slower than those of the bulk of \emph{Kepler} red giants. These stars might be the evolved counterparts of $\gamma$ Doradus-type stars that exhibit extremely slow near-core rotation during the main sequence (\citealt{LiGang2020binary}), which has been interpreted as a potential consequence of tidal interactions (\citealt{Fuller2021_inverse_tides}).}
%implying an extremely slow core rotation in progenitor stars. 

\sebnew{As a by-product of this work, }we also find a group of new stars which show smaller $\Delta\Pi_1$ compared to the $\Delta\Pi_1$--$\Delta\nu$ degenerate sequence. \LG{We confirm the previous discovery by \cite{Kuszlewicz2023ApJ} that the stars below the $\Delta\Pi_1$--$\Delta\nu$ degenerate sequence show smaller $q$ values, which is proof that stars might have undergone a process of mass transfer or binary merger.} We also discover that many stars show larger $\varepsilon_\mathrm{g}$, which might be a new clue for their central magnetic fields \citep{Li2022Nature}. \sebnew{We are currently investigating this possible interpretation.}

\begin{acknowledgements}
The authors acknowledge support from the project BEAMING ANR-18-CE31-0001 of the French National Research Agency (ANR) and from the Centre National d’Etudes Spatiales (CNES). 
\LG{The research leading to these results has received funding
from the KU\,Leuven Research Council (grant C16/18/005: PARADISE). 
GL acknowledges the Research Foundation Flanders (FWO) Grant for a long stay abroad 
(grant V422323N) and the Dick Hunstead Fund for Astrophysics for his 2-month stay at the University of Sydney.} 
GL acknowledge the travel support from the National Natural Science Foundation of China (NSFC) through the grant 12273002. 
GL thanks Dr. Yaguang Li for generously sharing their data. \LGvtwo{GL also thanks Professor Dennis Stello and Emily Hatt for the private discussion about the over-density ridge.}
This paper includes data collected by the \emph{Kepler} mission and obtained from the MAST data archive at the Space Telescope Science Institute (STScI). Funding for the \emph{Kepler} mission is provided by the NASA Science Mission Directorate. STScI is operated by the Association of Universities for Research in Astronomy, Inc., under NASA contract NAS 5–26555.
\end{acknowledgements}

% - use BibTeX with the regular commands:
   \bibliographystyle{aa} % style aa.bst
   \bibliography{main_paper} % your references Yourfile.bib

\begin{appendix}

\section{Observed splitting asymmetries in KIC\,11245496}\label{appendix_sec:read_asymmetries_in_11245496}

\begin{figure}
    \centering
    \includegraphics[width=\linewidth]{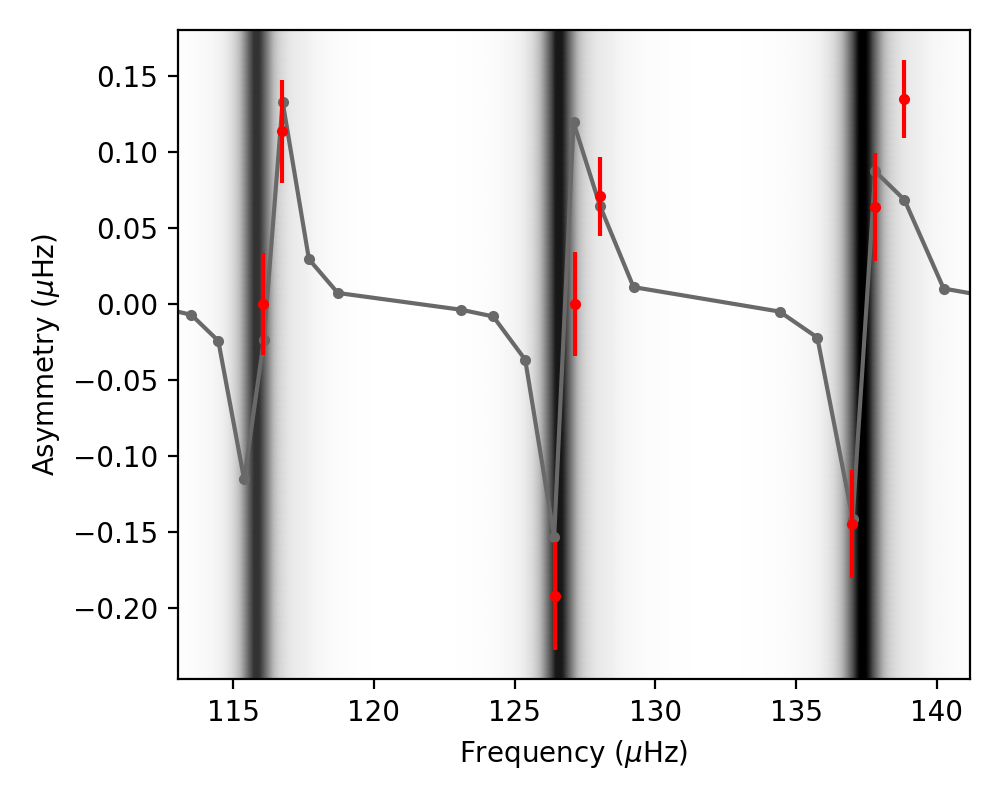}
    \caption{Observed asymmetries in $l=1$ rotational multiplets in KIC\,11245496. The difference to Fig.~\ref{fig_asym_NDE} is that the red dots are real observed values of asymmetries, so we can show error bars on them. }
    \label{fig_asym_NDE_real}
\end{figure}

\LGvtwo{In Fig.~\ref{fig_asym_NDE_real}, we present the observed splitting asymmetries for KIC\,11245496 as a comparison to the simulated splitting asymmetries shown in Fig.~\ref{fig_asym_NDE}. We find that $l=1$ modes exhibit observable splitting asymmetries similar to the $l=2$ modes reported by \citep{Deheuvels2017}. Moreover, Approach Two successfully reproduces the observed asymmetries.}

\section{The other parameters}\label{appendix_sec:other_parameters}

In this section, we show the other parameters that we measure, which may not be directly related to our scientific goals but still have some interesting results.

\subsection{shift of pure $l=1$ p mode $f_\mathrm{shift}$}

\begin{figure}
        \centering
        \includegraphics[width=\linewidth]{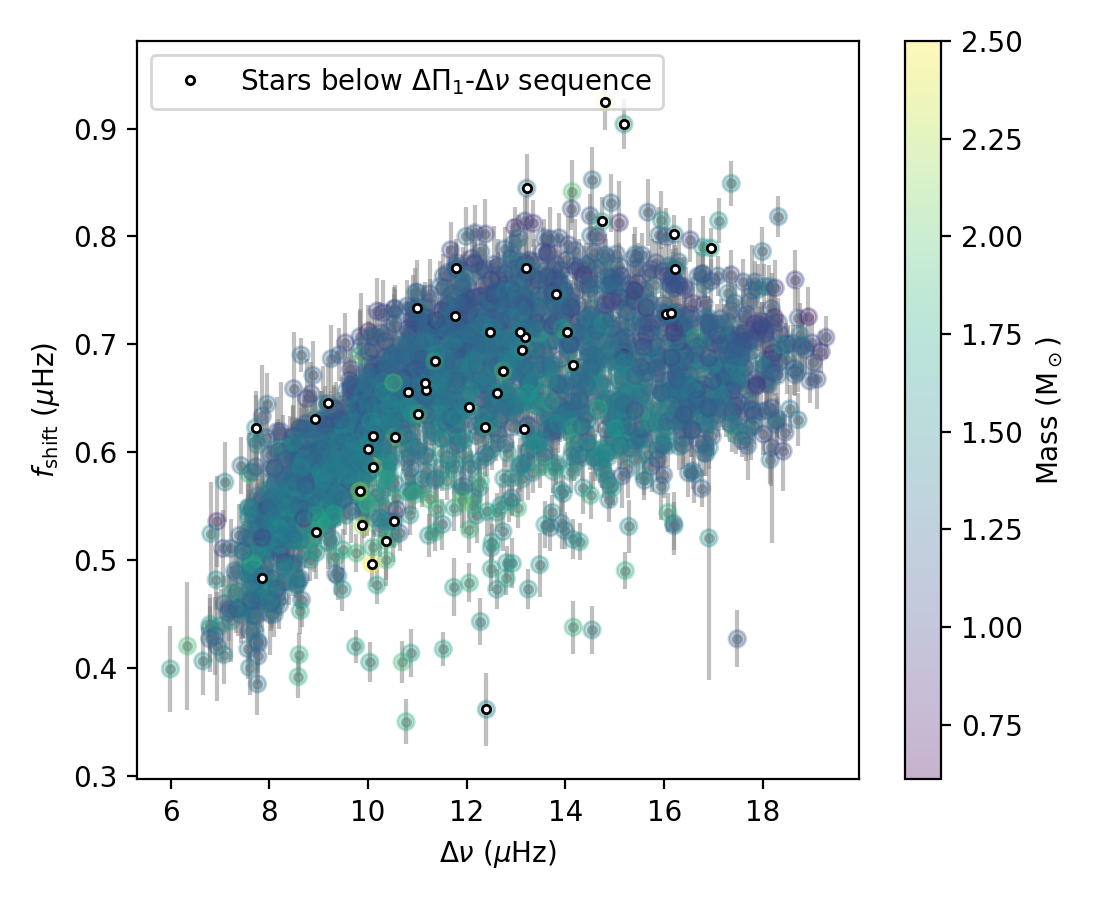}
        \caption{The shift of the pure $l=1$ p modes $f_\mathrm{shift}$ as a function of $\Delta \nu$. }
        \label{fig:p_mode_shift}
    \end{figure}

In eq.~\ref{eq:l_1_p_mode_asymptotic_relation}, we define the shift of the pure $l=1$ p modes, denoted as $f_\mathrm{shift}$, to properly account for the locations of invisible pure p modes. Figure~\ref{fig:p_mode_shift} illustrates a correlation between $f_\mathrm{shift}$ and $\Delta\nu$. As $\Delta\nu$ decreases, $f_\mathrm{shift}$ also decreases, ranging from approximately 0.8 to 0.4, and the spread of values becomes smaller. We observe a mass gradient in this relation, where higher-mass stars tend to exhibit smaller values of $f_\mathrm{shift}$.

\begin{figure}
    \centering
    \includegraphics[width=\linewidth]{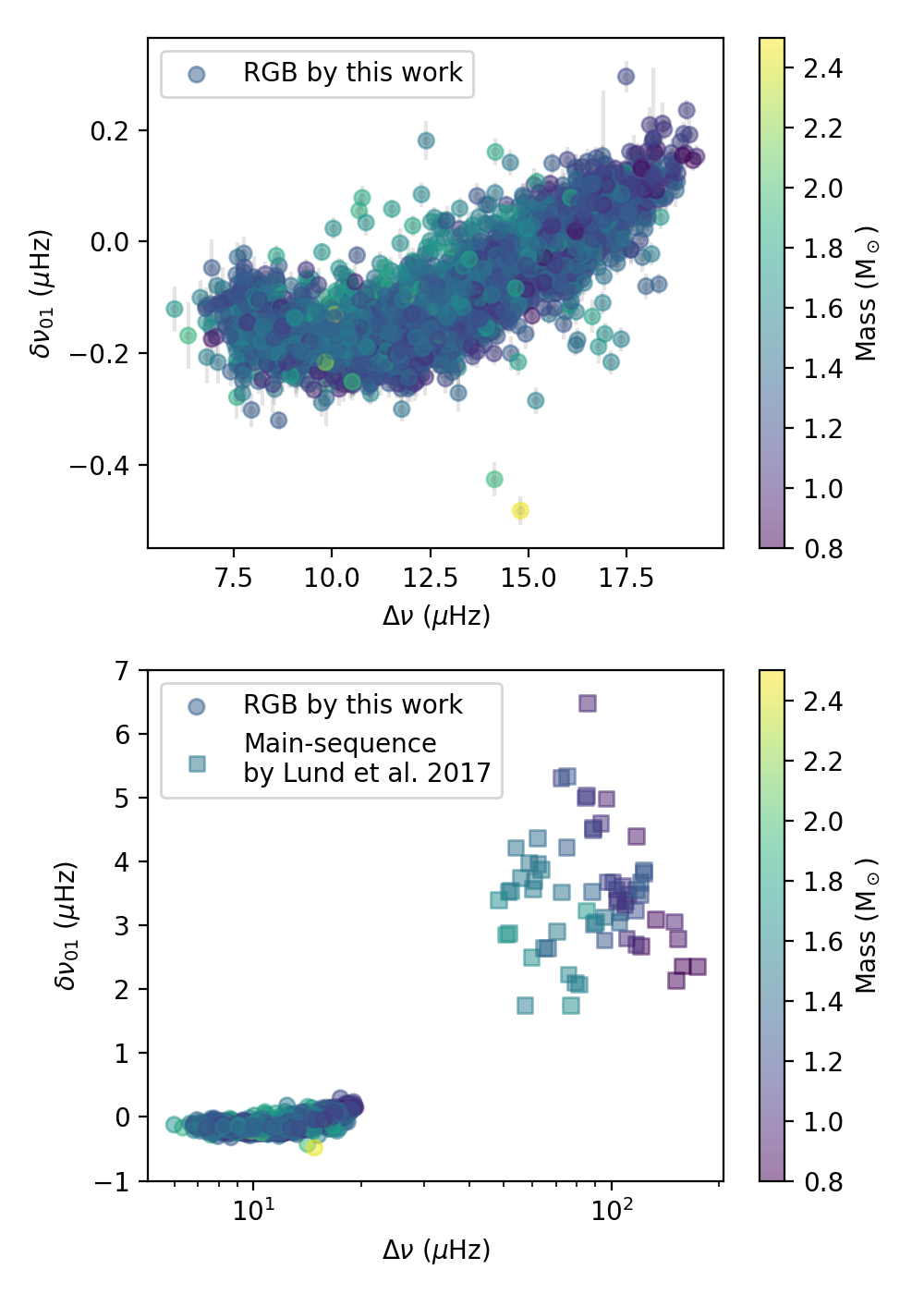}
    \caption{The small separation $\delta_{01}$ with $\Delta\nu$. Top panel: the RGB stars by this work. Bottom panel: both the RGB stars by this work and the main-sequence stars by \cite{Lund2017ApJ}. }
    \label{fig:delta_nu_01}
\end{figure}

Therefore, we can measure the small separation $\delta \nu_{01}$ defined as $\delta \nu_{01}\equiv\frac{1}{2}\left(\nu_{n, 0}+\nu_{n+1, 0}\right)-\nu_{n, 1}$. The top panel of Fig.~\ref{fig:delta_nu_01} shows $\delta \nu_{01}$ of the RGB stars by this work. $\delta \nu_{01}$ decreases from $\sim 0.2\,\mathrm{\mu Hz}$ to $\sim -0.2\,\mathrm{\mu Hz}$ as $\Delta \nu$ decreases, thus for RGB stars the pure $l=1$ p modes are almost exactly located at the halfway between $l=0$ modes. The bottom panel shows how $\delta\nu_{0,1}$ evolves from the main sequence to the red-giant branch. The main-sequence stars by \cite{Lund2017ApJ} show a relatively large spread of $\delta\nu_{01}$, ranging from $2\,\mathrm{\mu Hz}$ to $6\,\mathrm{\mu Hz}$, and their $\delta\nu_{01}$ decreases with evolution and eventually converges to the RGB stars. 

\subsection{Inclination $i$}
\begin{figure}
    \centering
    \includegraphics[width=\linewidth]{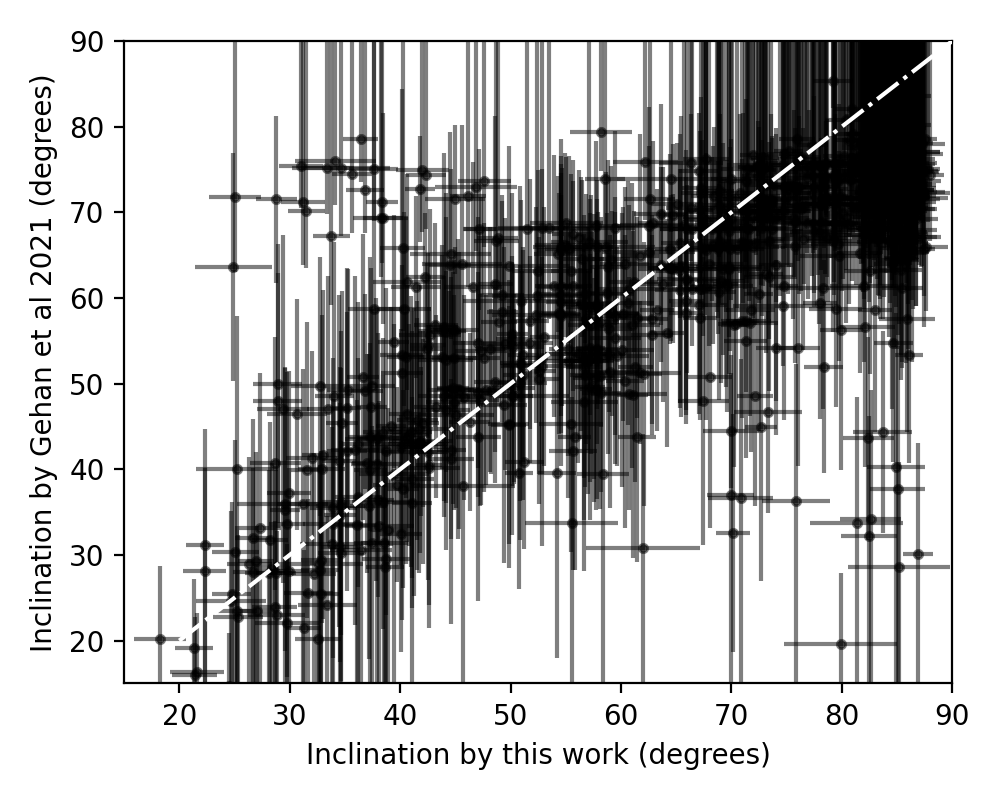}
    \caption{Comparison between the inclinations measured by this work and those reported by \cite{Gehan2021}. The white dotted-dashed line stands for the 1:1 relation.}
    \label{fig:comparison_inclination}
\end{figure}

We obtain the stellar inclinations when fitting the Lorentzian profiles to the rotational splittings. Figure~\ref{fig:comparison_inclination} compares our results with those reported by \cite{Gehan2021}. Our results show a general consistency with the inclinations by \cite{Gehan2018}. The difference might come from the different mode identifications or different fitting strategies. For example, we fit the Lorentzian profiles to all the rotational splittings (as shown in Fig.~\ref{fig:whole_spectrum_fit_of_8636389}), whereas \cite{Gehan2021} only used g-dominated modes to fit. 

\begin{figure}
    \centering
    \includegraphics[width=\linewidth]{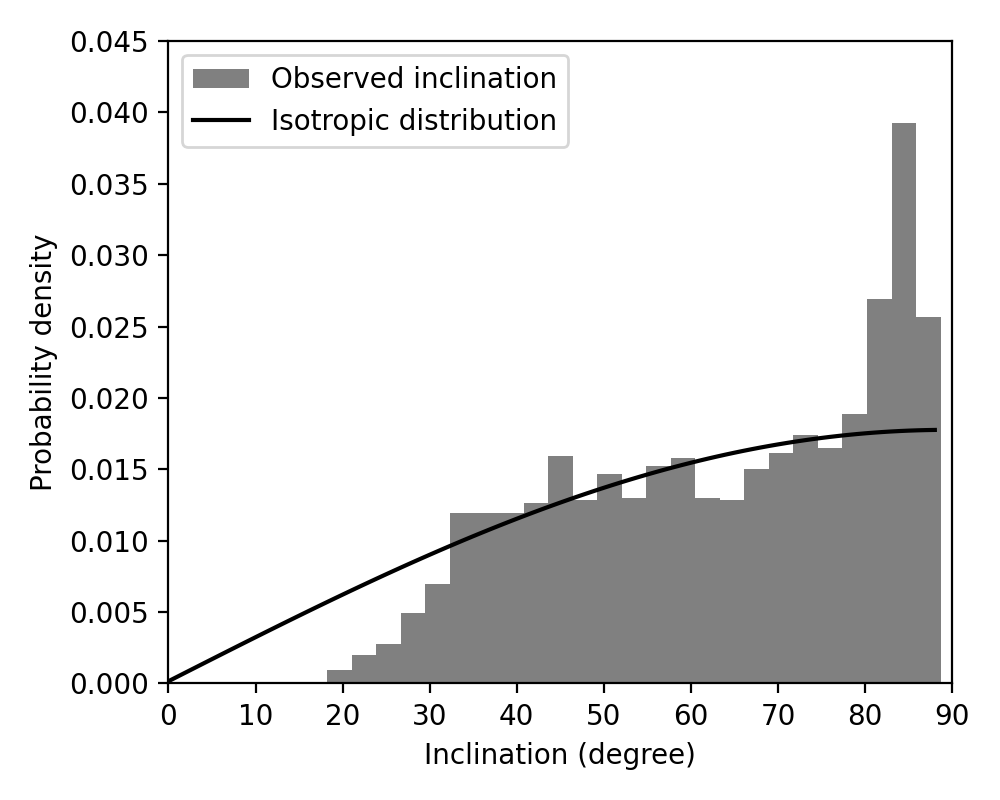}
    \caption{The distribution of stellar inclination measured by rotational splittings. }
    \label{fig:inclination_distribution}
\end{figure}

Figure~\ref{fig:inclination_distribution} displays the distribution of stellar inclinations in our sample. The solid curve is the theoretical distribution that is proportional to $\sin (i)$, assuming isotropic orientations of stellar rotational axes. Similar to the findings of \cite{Gehan2021}, we also observe a distribution that deviates from the isotropic expectation. In fig.~\ref{fig:inclination_distribution}, the number increases rapidly from $i=20^\circ$ to $30^\circ$ because the stars with smaller inclinations do not show any rotational splittings hence cannot be measured in our algorithm. Additionally, we identify an excess when the inclination is close to $90^\circ$, which has been referred to as "fit locking" and explained by \cite{Ballot2008}. Therefore, the deviation from isotropic distribution cannot prove that the directions of the stellar rotational axes are not isotropic.

\section{The spectrum fit of KIC\,8636389 \label{app_spectrum_8636389}}

\begin{figure*}
    \centering
    \includegraphics[width=\linewidth]{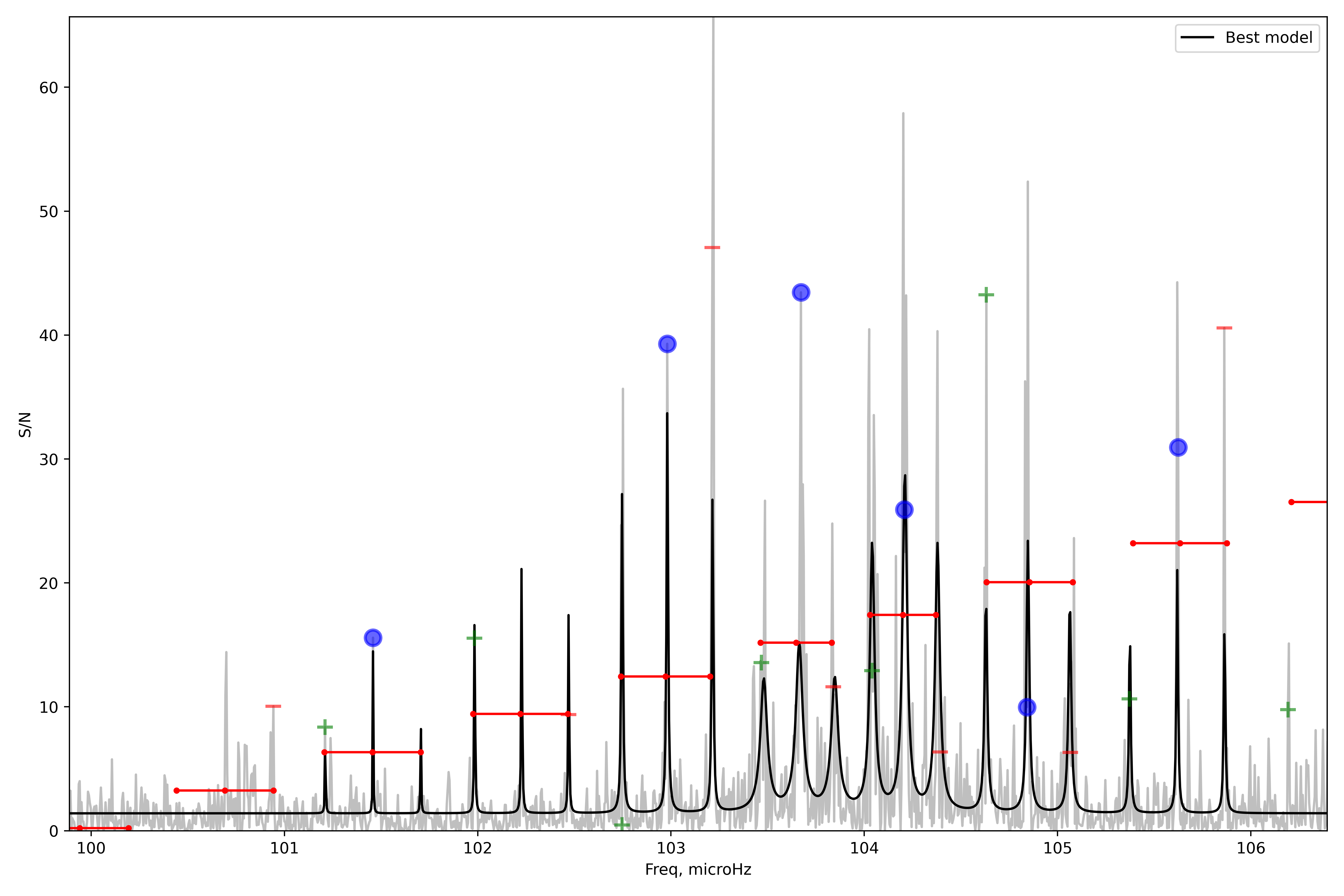}
    \caption{The fit of the whole power spectrum of KIC\,8636389. We only show the sector with $n_\mathrm{p} = 10$. The grey line is the observed power spectrum. We mark the $m=-1$, $0$, and $+1$ modes by the red short line, blue circle, and green plus symbols. The horizontal red lines mark the rotational splittings. The best-fitting result is shown by the black line, from which the frequency, rotational splitting, amplitude, stellar inclination, and linewidth are derived. }
    \label{fig:whole_spectrum_fit_of_8636389}
\end{figure*}

We show the mode identifications of azimuthal order $m$ and Lorentzian fit of the mixed-mode rotational splittings of KIC\,8636389 in Fig.~\ref{fig:whole_spectrum_fit_of_8636389}. The spectrum in fig.~\ref{fig:whole_spectrum_fit_of_8636389} are located between $n_\mathrm{p}=10,l=0$ p mode and $n_\mathrm{p}=10,l=2$ p mode. The best-fitting model reproduces the observed power spectrum and gives the measurements of oscillation frequencies, linewidth changes and inclination. 

\section{The rotational splittings and the Lorentzian fit of KIC\,7630743}

\begin{figure}
    \centering
    \includegraphics[width=0.72\linewidth]{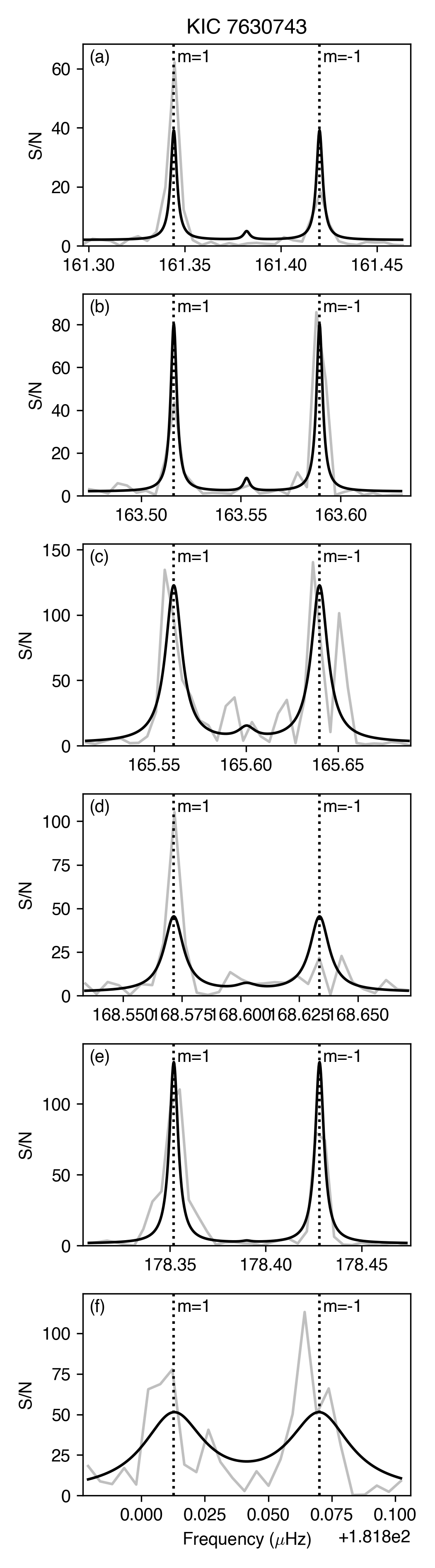}
    \caption{The rotational splittings of KIC\,7630743. the grey lines show the observed power spectrum and the solid black lines show the best-fitting Lorentzian profiles. The vertical dashed lines mark the locations of $m=\pm1$ modes.}
    \label{fig:splittings_of_7630743}
\end{figure}

In fig.~\ref{fig:splittings_of_7630743} we show the detail of the observed and fitting rotational splittings of KIC\,7630743. This star shows very slow rotational rates and a near-rigid radial rotation profile (see the discussion in section~\ref{subsec:differential_rotation}).  

%\section{visually inspection example for 2:1 rotation rate with Gehan}

\section{Estimating near-degeneracy-effects on dipole mixed modes in a \textit{Kepler} red giant \label{app_NDE}}

To estimate NDE in the chosen reference star, we followed \cite{Deheuvels2017}. The authors expressed the mode displacement as a linear combination of the eigenfunctions of the unperturbed degenerate modes. They focused on $l=2$ mixed modes, which have only weak coupling. Thus, they restricted their analysis to the case of two near-degenerate modes. For $l=1$ modes, the coupling between the cavities is stronger, so that one p mode couples to several consecutive g modes. When NDE are important, one should thus consider the general case of $N$ degenerate modes. Contrary to what was stated by \cite{ong22}, the development proposed in \cite{Deheuvels2017} can easily be extended to the case of $N$ modes. The eigenfunctions  of perturbed modes are then simply written as $\vect{\xi} = \sum_{i=0}^{N-1} c_i \vect{\xi}_{0,i}$, where $\vect{\xi}_{0,i}$ is the eigenfunction of the $i^{\rm th}$ unperturbed mode experiencing near-degeneracy. The frequencies of the near-degenerate modes are then found as the eigenvalues of the matrix $\mathsf{M} = \mathsf{A} + \mathsf{R}$, where $\mathsf{A}$ is a diagonal matrix containing the square of the unperturbed frequencies of the $N$ considered modes ($\mathsf{A}_{ii} = \omega_{0,i}^2$), and the components of matrix $\mathsf{R}$ are given by
\begin{equation}
    \mathsf{R}_{ij} = \langle \delta\mathcal{L}_{\rm R} \vect{\xi}_{0,i} , \vect{\xi}_{0,j} \rangle, \\
\end{equation}
where $\delta\mathcal{L}_{\rm R}$ is the rotational perturbation to the oscillation operator and $\langle \cdot , \cdot \rangle$ is the inner product defined as $\langle \vect{\xi} , \vect{\eta} \rangle = \int \vect{\xi}^* \cdot \vect{\eta} \,\hbox{d}m$ for functions $\vect{\xi}$ and $\vect{\eta}$. To characterise $\delta\mathcal{L}_{\rm R}$, we needed to assume a rotation profile $\Omega(r)$. We chose a two-zone profile with a discontinuity in the evanescent zone. We then solved the eigenvalue problem to obtain the perturbed mode frequencies of the considered stellar models.

\end{appendix}
   
\end{document}